\documentclass[12pt,a4paper]{article}
\usepackage{amsmath}
\usepackage{amsxtra}
    \usepackage{amstext}
    \usepackage{amssymb}
    \usepackage{latexsym}
    \usepackage{graphicx}
\usepackage{color}
\usepackage{graphics}
\usepackage{float}
\usepackage{wrapfig}
\usepackage{cite}

\newcommand{\p}{\vspace{6pt}\noindent}
\newcommand{\jump}{\vspace{2pt}}

\advance\textheight by \topskip
\makeatletter

\@addtoreset{equation}{section}
\def\section{\@startsection {section}{1}{\z@}{-8.5ex plus -1ex minus
 -.2ex}{3.3ex plus .2ex}{\large\bf}}
\def\subsection{\@startsection{subsection}{2}{\z@}{-3.25ex plus
 -1ex minus -.2ex}{1.5ex plus .2ex}{\bf}}
\def\subsubsection{\@startsection{subsubsection}{3}{\z@}{-3.25ex plus%
 -1ex minus -.2ex}{1.5ex plus .2ex}{\sl}}

\begin{document}

\begin{titlepage}
\vspace*{-2cm}
\begin{flushright}

\end{flushright}

\vspace{0.3cm}

\begin{center}
{\Large {\bf Adding integrable defects to the Boussinesq equation}} \\
\vspace{1cm} {\large  E.\ Corrigan\footnote{\noindent E-mail: {\tt edward.corrigan@york.ac.uk}}}\\
\vspace{0.5cm}
{\em Department of Mathematics \\ University of York, York YO10 5DD, U.K.} \\
\vspace{0.3cm} {\large and}\\ \vspace{0.5cm}
{\large C.\ Zambon\footnote{\noindent E-mail: {\tt cristina.zambon@durham.ac.uk}}} \\
\vspace{0.3cm}
{\em Department of Physics \\ Durham University, Durham DH1 3LE, U.K.} \\

\vspace{2cm} {\bf{ABSTRACT}}\\

\end{center}    \vspace{.5cm}
\narrower{\p The purpose of this paper is to extend the store of models able to support integrable defects by investigating the two-dimensional Boussinesq nonlinear wave equation. As has been previously noted in many examples, insisting that a defect contributes to energy and momentum to ensure their conservation, despite the presence of discontinuities and the explicit breaking of translation invariance, leads to sewing conditions relating the two fields and their derivatives on either side of the defect. The manner in which several types of soliton solutions to the Boussinesq equation are affected by the defect is explored and reveals new effects that have not been observed in other integrable systems, such as the possibility of a soliton reflecting from a defect or of a defect decaying into one or two solitons.}

\p \\

\vfill
\end{titlepage}

\section{Introduction}\label{intro}

The Boussinesq equation was introduced 150 years ago to provide an approximate description of water waves \cite{b1872}. Since then, many interesting features of this equation have been discovered but it would be difficult to review them all here, or even to provide a comprehensive list of references. For this article, the papers  \cite{h1973,hs1977,as1978, h1982, lmk1987, mos1988, bz2002, rs2017} have been particularly useful.
The purpose of this paper, motivated by the relevance of the Boussinesq equation, or more accurately a perturbed Boussinesq equation, to the propagation of solitons along nerve fibres \cite{katz1966,heimberg2005, lautrup2011, appali2012}, is to examine the possibility of constructing \lq integrable discontinuities', or  \lq defects', as a first tentative step towards attempting to model a synapse \cite{katz1966} as a field discontinuity. If that is possible, a synapse would be described by sewing conditions relating the (perturbed-)Boussinesq fields on either side of it. Since it turns out that in all examples investigated so far integrable defects are purely transmitting, at least until now, and therefore efficient from the point of view of processing solitons, this is a property one could hope a synapse would have, at least approximately.

\p Integrable defects  (sometimes called \lq jump defects' to distinguish them from other types of impurity) have been studied for some time in a variety of two-dimensional (one space - one time) contexts. The study of impurities has a long history in solid-state physics but the lines of thought relevant to the discussion presented here really began with papers such as \cite{dms1994,kl1999}. On the other hand, Lagrangian descriptions of integrable defects were introduced more recently \cite{bcz2003}, and led to discoveries within the sine-Gordon model \cite{bcz2003, bcz2005, hk2007, n2009, cz2010, ad2012},   the KdV equation \cite{cz2005}, in the nonlinear Schr\"odinger equation \cite{cz2005,c2008,ad2012nls,z2014}, and in conformal and affine Toda field theories \cite{bcz2004, cz2007, cz2009a,cz2009,cr2013,br2017}. Other examples demonstrating various features of integrable defects, though  less relevant to the present context, can be found in the collection of papers  \cite{gyz2006,d2016,cp2017,pd2019,cz2020}.

\p In all examples studied so far there is a principal recurring theme: an integrable defect is special because its defining property is the capacity to make a contribution to both energy and momentum to ensure that the total energy and momentum, meaning the contributions from fields on either side of the defect together with the defect contribution, is conserved. It is unsurprising that the total energy should be conserved because a defect of the type considered here does not break time translation invariance. However, a defect placed at a specific point on the spatial axis manifestly breaks translation invariance and therefore it is surprising the defect can compensate for this. In fact, in all situations studied so far, the insistence on a conserved overall momentum including a defect contribution actually requires the fields to either side of a defect to satisfy integrable field equations.

\p One interesting feature of \lq integrable defects' is the fact they are essentially purely transmitting. This means a soliton encountering a defect, for example within the sine-Gordon model, will pass through the defect suffering a delay, or will convert to an anti-soliton with a delay, or will be absorbed by the defect. The three  different possibilities occur according to whether the rapidity of the soliton is less than, greater than, or exactly equal to, the (positive) free parameter associated with the defect \cite{bcz2005}. If it proves to be possible to construct a similar defect within the Boussinesq setup then it will be interesting to see precisely how solitons behave as they scatter with it. The purpose of this paper is to demonstrate that defects of this type are indeed supported by the Boussinesq equation and that solitons are typically transmitted by the defect rather than being reflected or destroyed by it. However, there are particular special solutions to the Boussinesq equation that demonstrate features not shared with sine-Gordon solitons. Examples of these are \lq merging or splitting' soliton solutions \cite{rs2017}, which do not preserve the number of solitons as they evolve, and it transpires these can exhibit previously unseen novel behaviour when encountering a defect.

\p The plan of the paper is as follows. In section 2, the starting point and conventions to be used will be established. In section 3, the constraints following from insisting that there be a conserved total energy and momentum, including contributions from the defect, are analysed in detail leading to explicit expressions for the necessary sewing conditions. In section 4, the linear approximation is analysed to demonstrate the purely transmitting property in the simplest case of a monochromatic wave. In section 5, the scattering of one or two soliton by the defect is analysed and then compared with two-soliton scattering. For these, the defect is purely transmitting. As mentioned above, the Boussinesq equation also permits solutions that allow the soliton number to change in finite time. The latter are also discussed in section 5 and lead to a wider range of behaviours when encountering a defect. Finally, section 6 contains some concluding remarks concerning a soliton solution to the perturbed Boussinesq equation studied by the authors of \cite{heimberg2005,lautrup2011,appali2012}.

\section{The Boussinesq equation}

A convenient starting point for the purposes of this article is the Lagrangian density provided in ref\cite{bz2002}:
\begin{equation}\label{BZlagrangian}
  {\cal L}=\frac{1}{2c^2}\,u_t^2 -\frac{1}{2} \,u_x^2 -\frac{\lambda}{3}\,u_x^3 -\frac{\kappa^2}{2}\,u_{xx}^2.
\end{equation}
Then, the Euler-Lagrange equations require that $u$ satisifies the nonlinear wave equation:
  \begin{equation}\label{BZueq}
  \frac{1}{c^2}\, u_{tt}-u_{xx}-\lambda\left(u_x^2\right)_x +\kappa^2 u_{xxxx}=0.
  \end{equation}
  Differentiating this with respect to $x$ and letting $w=u_x$, the field $w$ satisifies:
  \begin{equation}\label{BZweq}
  \frac{1}{c^2}\, w_{tt}-w_{xx}-\lambda\left(w^2\right)_{xx} +\kappa^2 w_{xxxx}=0,
  \end{equation}
  which is the Boussinesq equation \cite{b1872}. Here, dimensional constants $c,\ \lambda,\ \kappa$ are inserted explicitly, where $c$ is a speed, $\kappa$ is a length and $\lambda$ is a length divided by the dimension of $u$. Note, the sign of $\lambda$ is unimportant since changing the sign of $\lambda$ is equivalent to changing the sign of $u$.

\p  In terms of $u$, the energy density associated with \eqref{BZlagrangian} is:
  \begin{equation}\label{EB}
    {\cal E}=\frac{1}{2c^2}\,u_t^2 +\frac{1}{2} \,u_x^2 +\frac{\lambda}{3}\,u_x^3 +\frac{\kappa^2}{2}\,u_{xx}^2,
  \end{equation}
  and the momentum density is
  \begin{equation}\label{PB}
    {\cal P}=-\frac{1}{c^2}\,u_t\, u_x.
  \end{equation}
  The minus sign is included in \eqref{PB} to ensure a soliton travelling along the $x$-axis in the sense of increasing $x$ has positive momentum. It is also useful to note for future reference, using equation \eqref{BZueq}, that the time derivatives of the energy and momentum densities satisfy:
\begin{equation}\label{Et}
{\cal E}_t=\frac{\partial}{\partial x}\left(u_t(u_x+\lambda u_x^2-\kappa^2u_{xxx})+\kappa^2u_{xx}u_{xt}\right),
\end{equation}
and
\begin{equation}\label{Pt}
{\cal P}_t=-\frac{\partial}{\partial x}\left(\frac{1}{2c^2}u_t^2+\frac{1}{2}u_x^2+\frac{2\lambda}{3}\,u_x^3-\kappa^2u_xu_{xxx}+\frac{\kappa^2}{2}\, u_{xx}^2\right)\,.
\end{equation}
This is the formulation mainly used in this article. Also, throughout it will be assumed that $\kappa$ is real and positive, which is the form of the Boussinesq equation that is most relevant to the investigation of pulse propagation in nerve fibres \cite{heimberg2005}.
Sometimes, with $\kappa^2>0$, \eqref{BZweq} is called the \lq good'  Boussinesq equation.

\p For future use, the equation \eqref{BZueq} has the well-known soliton solution \cite{h1973,hs1977,mos1988} progressing with speed $\nu$:
\begin{equation}\label{Bsoliton}
  u(x,t)=-\frac{3\kappa\gamma(\nu)}{\lambda}\, \tanh\left(\frac{\gamma(\nu)}{2\kappa}\,(x-\nu t)\right),\ \ \gamma(\nu)=\sqrt{1-\nu^2/c^2},\ \ |\nu|<c.
\end{equation}
A direct calculation of the energy and momentum for the soliton, using the above expressions  \eqref{EB} and \eqref{PB} for the energy and momentum densities, gives
\begin{equation}\label{EPsoliton}
P=M_0\gamma^3 \nu,\ \ E=\frac{1}{5}\,M_0 c^2\gamma^3 \, \left(1+\frac{4 \nu^2}{c^2}\right),\ \ \hbox{where} \ \ M_0=\frac{6\kappa}{\lambda^2 c^2}.
\end{equation}
These expressions look strange. However, if $\nu<< c$, they are approximated by:
\begin{equation}\label{EPslowsoliton}
P=M_0\gamma^3 \nu\approx M_0\,\nu,\ \ E=\frac{1}{5}\,M_0 c^2 \gamma^3 \, \left(1+\frac{4 \nu^2}{c^2}\right)\approx \frac{1}{5}\, M_0\, c^2 +\frac{1}{2}\, M_0\, \nu^2,
\end{equation}
 which look more familiar. On the other hand, as $\nu\rightarrow c$, both $P$ and $E$ tend towards zero, as they should, since according to eq\eqref{Bsoliton} $u\rightarrow 0$ in that limit and the soliton disappears.

\p If $\kappa^2<0$, then the corresponding soliton solution would be real provided $|\nu|>c$. This illustrates the distinction between the \lq good' and \lq bad' variants of the equation.

\section{Adding an integrable defect}

As in previously discussed models, an \lq integrable' defect can be added to the Boussinesq setup at the point $x=x_0$ by considering two different Boussinesq fields (by default these are now described by eq\eqref{BZueq} with $\kappa^2>0$), $u$ defined in the region $x<x_0$ and $v$ defined in the region $x>x_0$, together with sewing conditions that relate the fields and their derivatives across the defect at $x=x_0$. Note, in this context fields and their derivatives are defined as limits from the left of $x_0$ (for $u$) and from the right of $x_0$ (for $v$). Typically, a defect will require discontinuities in the fields and their derivatives at $x=x_0$; in particular, there is no assumption that $u(x_0,t)=v(x_0,t)$.
For these reasons, the action will be taken to be
\begin{equation}\label{defectaction}
{\cal A}=\int dt\left[\int_{-\infty}^{x_0}dx\, {\cal L}_u +{\cal D}+\int_{x_0}^\infty dx\, {\cal L}_v\right],
\end{equation}
where ${\cal L}_u,\ {\cal L}_v$ are both of the form \eqref{BZlagrangian}, and ${\cal D}$ depends on the fields $u,v$ and their space/time derivatives evaluated at $x=x_0$ (in the sense described above).

\p Note also, it is implicitly assumed the Boussinesq equations in the regions $x<x_0$ and $x> x_0$ are identical. However, there is no need to make this assumption since the media to either side of the defect could, in principle, be different, meaning the dimensional constants in eq\eqref{BZueq} might be different. An example of a change of medium of this kind, for the sine-Gordon model, is given in \cite{cz2020}. In such a situation, a defect involving a discontinuity of the fields and/or their derivatives is to be expected.

\p Starting with the action given in eq\eqref{defectaction} and varying the fields $u,v$ independently in their respective domains leads to the Boussinesq equations for $u,v$ within their respective domains, together with the sewing conditions that should hold at $x=x_0$:
\begin{eqnarray}
 \nonumber \frac{\partial{\cal L}_u}{\partial u_x}-\frac{\partial}{\partial x}\,\frac{\partial{\cal L}_u}{\partial u_{xx}}+\frac{\partial{\cal D}}{\partial u}-\frac{\partial}{\partial t}\,\frac{\partial{\cal D}}{\partial u_t} &=& 0,  \\ \nonumber
\frac{\partial{\cal L}_u}{\partial u_{xx}}+\frac{\partial{\cal D}}{\partial u_x}&=&0,  \\ \nonumber
-\frac{\partial{\cal L}_v}{\partial v_x}+\frac{\partial}{\partial x}\,\frac{\partial{\cal L}_v}{\partial v_{xx}}+\frac{\partial{\cal D}}{\partial v}-\frac{\partial}{\partial t}\,\frac{\partial{\cal D}}{\partial v_t} &=& 0,\\
  - \frac{\partial{\cal L}_v}{\partial v_{xx}}+\frac{\partial{\cal D}}{\partial v_x}&=&0.
\end{eqnarray}
Using the explicit expressions for the Lagrangians these become conditions on the space derivatives of $u$ and $v$ at $x=x_0$:
\begin{eqnarray}\label{uvconditions} \nonumber
\frac{\partial{\cal L}_u}{\partial u_x}-\frac{\partial}{\partial x}\,\frac{\partial{\cal L}_u}{\partial u_{xx}}&=& -u_x-\lambda u_x^2 +\kappa^2 u_{xxx}=-\frac{\partial{\cal D}}{\partial u}+\frac{\partial}{\partial t}\,\frac{\partial{\cal D}}{\partial u_t},\\ \nonumber
\frac{\partial{\cal L}_u}{\partial u_{xx}}&=&-\kappa^2u_{xx}=-\frac{\partial{\cal D}}{\partial u_x},\\ \nonumber
\frac{\partial{\cal L}_v}{\partial v_x}-\frac{\partial}{\partial x}\,\frac{\partial{\cal L}_v}{\partial v_{xx}}&=& -v_x-\lambda v_x^2 +\kappa^2 v_{xxx}=\frac{\partial{\cal D}}{\partial v}-\frac{\partial}{\partial t}\,\frac{\partial{\cal D}}{\partial v_t},\\
-\frac{\partial{\cal L}_v}{\partial u_{xx}}&=&\kappa^2v_{xx}=-\frac{\partial{\cal D}}{\partial v_x}\,.
\end{eqnarray}
Note: these expressions suggest strongly that ${\cal D}$ depends only on $u,v,u_x, v_x,u_t,v_t$ (all evaluated at $x_0$ via appropriate limits). These will be used to replace the second and third order space derivatives evaluated at $x_0$ whenever they occur.

\p The total energy contributed by the fields to either side of the defect is given by
\begin{equation}
E = E_u +E_v =\int_{-\infty}^{x_0}dx\, {\cal E}_u +\int_{x_0}^\infty dx\, {\cal E}_v.
\end{equation}
This is not necessarily conserved but, using the expression \eqref{Et}, and assuming the fields and their derivatives are decaying to zero as $x\rightarrow \pm\infty$, its time derivative can be expressed as the difference of contributions from the defect at $x=x_0$. Thus,
\begin{equation}
E_t=\left.\left(u_t(u_x+\lambda u_x^2-\kappa^2u_{xxx})+\kappa^2u_{xx}u_{xt}\right)\right|_{x_0}-\left.\left(v_t(v_x+\lambda v_x^2-\kappa^2v_{xxx})+\kappa^2v_{xx}v_{xt}\right)\right|_{x_0},
\end{equation}
which can be rewritten using the sewing conditions \eqref{uvconditions} to obtain:
\begin{equation}\label{}
 E_t= \frac{\partial}{\partial t}\left({\cal D}-u_t\frac{\partial{\cal D}}{\partial u_t}-v_t\frac{\partial{\cal D}}{\partial v_t}\right).
\end{equation}
Hence, the quantity
\begin{equation}\label{Etotal}
E_{\rm tot}=E+u_t\frac{\partial{\cal D}}{\partial u_t}+v_t\frac{\partial{\cal D}}{\partial v_t}-{\cal D}\equiv E_u+E_v+E_{\cal D}
\end{equation}
is conserved independently of the specific details of the defect. This was to  be expected since the defect does not destroy the property of time translation invariance; it simply contributes to the total energy as shown.

\p On the other hand, placing a defect at a specified location certainly breaks translation invariance. Nevertheless, as has been demonstrated several times before in a variety of other models, insisting that a suitably modified total momentum is also conserved is a powerful constraint with interesting consequences. Moreover, evidence from examples examined so far suggests this is only possible in an integrable situation. Since the Boussinesq equation is known to be integrable it is worth exploring in detail the restrictions on the defect contribution ${\cal D}$ to the action.

\p The combined contributions from the fields to the momentum is
\begin{equation}
P = P_u +P_v =\int_{-\infty}^{x_0}dx\, {\cal P}_u +\int_{x_0}^\infty dx\, {\cal P}_v.
\end{equation}

\p Using the expression \eqref{Pt}  the time derivative of the total momentum is given by
\begin{eqnarray}\label{} \nonumber
 - P_t&=&\left.\left(\frac{1}{2c^2}u_t^2+\frac{1}{2}u_x^2+\frac{2\lambda}{3}\,u_x^3-\kappa^2u_xu_{xxx}+\frac{\kappa^2}{2}\, u_{xx}^2\right)\right|_{x_0}\\ \nonumber
&&\phantom{mmmmmm}-\left.\left(\frac{1}{2c^2}v_t^2+\frac{1}{2}v_x^2+\frac{2\lambda}{3}\,v_x^3-\kappa^2v_xv_{xxx}+\frac{\kappa^2}{2}\, v_{xx}^2\right)\right|_{x_0},
\end{eqnarray}
which can be rewritten, using the sewing conditions to replace the second and third derivative terms, to obtain
\begin{eqnarray}\label{} \nonumber
 - P_t&=&\left.\left(\frac{1}{2c^2}u_t^2-\frac{1}{2}u_x^2-\frac{\lambda}{3}\,u_x^3+\frac{1}{2\kappa^2}\left(\frac{\partial{\cal D}}{\partial u_x}\right)^2+u_x \frac{\partial{\cal D}}{\partial u}-u_x\frac{\partial}{\partial t}\,\frac{\partial{\cal D}}{\partial u_t}\right)\right|_{x_0}\\
&&\phantom{mmmm}-\left.\left(\frac{1}{2c^2}v_t^2-\frac{1}{2}v_x^2-\frac{\lambda}{3}\,v_x^3+\frac{1}{2\kappa^2}\left(\frac{\partial{\cal D}}{\partial v_x}\right)^2-v_x \frac{\partial{\cal D}}{\partial v}+v_x\frac{\partial}{\partial t}\,\frac{\partial{\cal D}}{\partial v_t}\right)\right|_{x_0}.
\end{eqnarray}
This expression can be rearranged slightly to give
\begin{eqnarray}\label{Ptrearranged}
\nonumber
 && \left(-P+u_x\frac{\partial{\cal D}}{\partial u_t}+v_x\frac{\partial{\cal D}}{\partial v_t}\right)_t =\\
&&\phantom{mmmm}   \left.\left(\frac{1}{2c^2}u_t^2-\frac{1}{2}u_x^2-\frac{\lambda}{3}\,u_x^3+\frac{1}{2\kappa^2}\left(\frac{\partial{\cal D}}{\partial u_x}\right)^2+u_{x} \frac{\partial{\cal D}}{\partial u}+u_{xt}\,\frac{\partial{\cal D}}{\partial u_t}\right)\right|_{x_0}\phantom{mmmmmmmmmm}\\ \nonumber
&&\phantom{mmmmmm}-\left.\left(\frac{1}{2c^2}v_t^2-\frac{1}{2}v_x^2-\frac{\lambda}{3}\,v_x^3+\frac{1}{2\kappa^2}
\left(\frac{\partial{\cal D}}{\partial v_x}\right)^2-v_x \frac{\partial{\cal D}}
{\partial v}-v_{xt}\,\frac{\partial{\cal D}}{\partial v_t}\right)\right|_{x_0}\equiv\frac{d P_{\rm D}}{dt}\,,
\end{eqnarray}
where ${\cal D}$ has to be carefully chosen so that the two terms central to  this expression conspire together to be the complete time derivative of a function $P_{\rm D}$ of the fields and their derivatives. If this is possible then the suitably adjusted momentum,
\begin{equation}\label{Ptotal}
P_{\rm tot}=P-u_x\frac{\partial{\cal D}}{\partial u_t}-v_x\frac{\partial{\cal D}}{\partial v_t}+P_{\rm D}\equiv P_u+P_v+P_{\cal D},
\end{equation}
which includes an explicit defect contribution, $P_{\cal D}$, will be conserved.

\p To begin the analysis of \eqref{Ptrearranged}, it is natural to assume the quantity ${\cal D}$ depends linearly on the time derivatives $u_t, \, v_t$, and has no dependence on higher order time derivatives. In other words, it is assumed the defect does not add dynamics of its own, meaning it is type I and not type II in the sense of ref\cite{cz2009}. With that assumption in place, the defect introduces no new dynamics. Then, a suitable ansatz is to take
\begin{equation}\label{Bstart}
  {\cal D}=\left. (u_t\, F +v_t\, G +H)\right|_{x=x_0},
\end{equation}
where $F,G,H$ depend only on the fields $u,u_x, v, v_x$ evaluated at $x=x_0$. Note, in analysing \eqref{Ptrearranged} all fields and their derivatives are evaluated at the defect  (in the limiting sense mentioned before), and this will be understood to be the case from now on. Using this ansatz, eq\eqref{Ptrearranged} becomes
\begin{eqnarray}\label{Ptwithansatz}
\nonumber
 && \left(-P+u_x\frac{\partial{\cal D}}{\partial u_t}+v_x\frac{\partial{\cal D}}{\partial v_t}\right)_t =u_{xt}\,F + v_{xt}\,G\\
&&\phantom{mmmm}   +\left.\left(\frac{1}{2c^2}u_t^2-\frac{1}{2}u_x^2-\frac{\lambda}{3}\,u_x^3+\frac{1}{2\kappa^2}
\left(\frac{\partial{\cal D}}{\partial u_x}\right)^2+u_{x} \frac{\partial{\cal D}}
{\partial u}\right)\right.\phantom{mmmmmmmm}\\ \nonumber
&&\phantom{mmmmmm}-\left.\left(\frac{1}{2c^2}v_t^2-\frac{1}{2}v_x^2-
\frac{\lambda}{3}\,v_x^3+\frac{1}{2\kappa^2}\left(\frac{\partial{\cal D}}{\partial v_x}\right)^2-v_x \frac{\partial{\cal D}}{\partial v}\right)\right.\,.
\end{eqnarray}
Suppose the right hand side of eq\eqref{Ptwithansatz} is the time derivative of $P_D$, the contribution the defect supplies to momentum. Then $P_D$ has to be a suitably chosen function of $u,u_x,v,v_x$, and it is already clear that
\begin{equation}\label{FGdef}
  \frac{\partial P_D}{\partial u_x}=F,\ \  \frac{\partial P_D}{\partial v_x}=G,
\end{equation}
implying the constraint:
\begin{equation}\label{FGconstraint1}
   \frac{\partial F}{\partial v_x}= \frac{\partial G}{\partial u_x}\,.
\end{equation}

\p Also, the other pieces of the right hand side of eq\eqref{Ptwithansatz} fall into three groups. First, there are a number of terms that are quadratic in time derivatives $u_t, v_t$; these must cancel among themselves. Second, there are terms that contain no time derivatives; these must also cancel among themselves. Third, there is a group of terms that are linear in time derivatives and these must combine in such a manner that eq\eqref{Ptwithansatz} can be rewritten
\begin{equation}\label{}
  \left(-P+u_x\frac{\partial{\cal D}}{\partial u_t}+v_x\frac{\partial{\cal D}}{\partial v_t}\right)_t =u_{xt}\,F + v_{xt}\,G +u_t \frac{\partial P_D}{\partial u} +v_t\frac{\partial P_D}{\partial v}\equiv \frac{\partial P_D}{\partial t} .
\end{equation}
Each set of terms will be considered separately in the first instance.
\subsection{Terms quadratic in time derivatives}

\p There are three terms that are quadratic in time derivatives, proportional to one of $u_t^2, v_t^2$ and $u_tv_t$, which must separately vanish. In detail, these are:
\begin{equation}\label{ut2}
u_t^2:\qquad \frac{1}{2c^2}+\frac{1}{2\kappa^2}\left[\left(\frac{\partial F}{\partial u_x}\right)^2-\left(\frac{\partial F}{\partial v_x}\right)^2\right]=0,
\end{equation}
\begin{equation}\label{vt2}
v_t^2:\qquad \frac{1}{2c^2}+\frac{1}{2\kappa^2}\left[\left(\frac{\partial G}{\partial v_x}\right)^2-\left(\frac{\partial G}{\partial u_x}\right)^2\right]=0,
\end{equation}
\begin{equation}\label{utvt}
\phantom{mmml}u_tv_t:\qquad\phantom{mmml} \frac{1}{\kappa^2}\left[\frac{\partial F}{\partial u_x}\,\frac{\partial G}{\partial u_x} -\frac{\partial F}{\partial v_x}\,\frac{\partial G}{\partial v_x}\right]=0.\phantom{mmmml}
\end{equation}
Using eq\eqref{FGconstraint1} together with eq\eqref{utvt} (and assuming the partial derivatives do not vanish) yields a second relation:
\begin{equation}\label{FGconstraint2}
   \frac{\partial F}{\partial u_x}= \frac{\partial G}{\partial v_x},
\end{equation}
in turn implying that eq\eqref{ut2} and eq\eqref{vt2} represent the same constraint. Combining eq\eqref{FGconstraint1} and eq\eqref{FGconstraint2} implies that $F$ and $G$ satisfy
\begin{equation}\label{FGeqs1}
  \frac{\partial^2 F}{\partial u_x^2}= \frac{\partial^2 F}{\partial v_x^2}, \quad  \frac{\partial^2 G}{\partial u_x^2}= \frac{\partial^2 G}{\partial v_x^2}.
\end{equation}
The only functions of $u_x, v_x$ compatible with eqs\eqref{FGeqs1} and satisfying the constraint \eqref{ut2} are linear, implying
\begin{equation}\label{FandG1}
  F=\alpha u_x +\beta v_x +f,\quad G=\beta u_x +\alpha v_x +g, \quad \beta^2-\alpha^2 = \frac{\kappa^2}{c^2}.
\end{equation}
In \eqref{FandG1}, the coefficients $\alpha, \beta$ and the terms $f,g$ are functions of $u,v$ only. Note also, the functions $f$ and $g$ are not unique since any change of the form
\begin{equation}\label{fggauge}
  f\rightarrow f^\prime = f+\frac{\partial\Omega}{\partial u},\quad g\rightarrow g^\prime = g+\frac{\partial\Omega}{\partial v},
\end{equation}
where $\Omega$ depends on $u$ and $v$ but not their derivatives, simply adds a total time derivative to ${\cal D}$, which is unimportant for the action.

\p From \eqref{FandG1} and \eqref{FGdef} it is already clear that $P_D$ is a quadratic function of $u_x,v_x$.

\subsection{Terms linear in time derivatives}

\p The next step is to extract and consider the terms in \eqref{Ptwithansatz} that are linear in $u_t, v_t$. Thus
\begin{eqnarray}
   &&  \frac{\partial P_D}{\partial u}=u_x\frac{\partial F}{\partial u}+v_x\frac{\partial F}{\partial v} +\frac{1}{\kappa^2}\left(\frac{\partial F}{\partial u_x}\,\frac{\partial H}{\partial u_x}-\frac{\partial F }{\partial v_x}\,\frac{\partial H}{\partial v_x}\right),\\
   &&  \frac{\partial P_D}{\partial v}=u_x\frac{\partial G}{\partial u}+v_x\frac{\partial G}{\partial v} +\frac{1}{\kappa^2}\left(\frac{\partial G}{\partial u_x}\,\frac{\partial H}{\partial u_x}- \frac{\partial G}{\partial v_x}\,\frac{\partial H}{\partial v_x}\right).
\end{eqnarray}
Using the expressions for $F, G$ from \eqref{FandG1} these may be rewritten as follows:
\begin{eqnarray}\label{PDuvdervs}
 \nonumber  &&  \frac{\partial P_D}{\partial u}=u_x^2\frac{\partial \alpha}{\partial u}+v_x^2\frac{\partial \beta}{\partial v}+u_xv_x\left(\frac{\partial\beta}{\partial u}+ \frac{\partial\alpha}{\partial v}\right)+u_x \frac{\partial f}{\partial u}+v_x \frac{\partial f}{\partial v}\\ \nonumber && \phantom{mmmmmm}+\frac{1}{\kappa^2}\left(\alpha\,\frac{\partial H}{\partial u_x}-\beta\,\frac{\partial H}{\partial v_x}\right),\\
 \nonumber  &&\frac{\partial P_D}{\partial v}=u_x^2\frac{\partial \beta}{\partial u}+v_x^2\frac{\partial \alpha}{\partial v}+u_xv_x\left(\frac{\partial\alpha}{\partial u}+ \frac{\partial\beta}{\partial v}\right)+u_x \frac{\partial g}{\partial u}+v_x \frac{\partial g}{\partial v}\\ && \phantom{mmmmmm}+\frac{1}{\kappa^2}\left(\beta\,\frac{\partial H}{\partial u_x}-\alpha\,\frac{\partial H}{\partial v_x}\right)  .
\end{eqnarray}
Clearly, the four partial derivatives of $P_D$ given in \eqref{FGdef} and \eqref{PDuvdervs} must be compatible and this requires several further contraints. To examine these, compare in turn the $(u,u_x),\ (v, v_x),\ (u,v_x)$ and $ (v,u_x)$ pairs of derivatives to obtain:
\begin{eqnarray}\label{2nddervrels}
 \nonumber && (u,u_x):\quad u_x\frac{\partial\alpha}{\partial u}+v_x\frac{\partial\alpha}{\partial v}+\frac{1}{\kappa^2}\left(\alpha\frac{\partial^2H}{\partial u_x^2}-\beta\frac{\partial^2 H}{\partial u_x \partial v_x}\right)=0  \\
\nonumber  &&(v,v_x):\quad u_x\frac{\partial\alpha}{\partial u}+v_x\frac{\partial\alpha}{\partial v}-\frac{1}{\kappa^2}\left(\alpha\frac{\partial^2H}{\partial v_x^2}-\beta\frac{\partial^2 H}{\partial u_x \partial v_x}\right)=0 \\
\nonumber  &&(u,v_x):\quad u_x\frac{\partial \alpha}{\partial v}+v_x\left(2\frac{\partial \beta}{\partial v}-\frac{\partial\alpha}{\partial u}\right)+\frac{\partial f}{\partial v}- \frac{\partial g}{\partial u} -\frac{1}{\kappa^2}\left(\beta\frac{\partial^2H}{\partial v_x^2}-\alpha\frac{\partial^2 H}{\partial u_x \partial v_x}\right)=0\phantom{mmmm}\\
  &&(v,u_x):\quad  u_x\left(2\frac{\partial\beta}{\partial u}-\frac{\partial\alpha}{\partial v}\right)+v_x\frac{\partial\alpha}{\partial u}-\frac{\partial f}{\partial v} +\frac{\partial g}{\partial u}+\frac{1}{\kappa^2}\left(\beta\frac{\partial^2H}{\partial u_x^2}-\alpha\frac{\partial^2 H}{\partial u_x \partial v_x}\right)=0.\phantom{mmmm}
\end{eqnarray}
Adding the first pair of relations in \eqref{2nddervrels} implies
\begin{equation}\label{firstpair}
  2 u_x\frac{\partial\alpha}{\partial u}+2v_x\frac{\partial\alpha}{\partial v}+\frac{\alpha}{\kappa^2}\left(\frac{\partial^2H}{\partial u_x^2}-\frac{\partial^2H}{\partial v_x^2}\right)=0,
\end{equation}
while adding the third and fourth relations gives
\begin{equation}\label{secondpair}
2 u_x\frac{\partial\beta}{\partial u}+2v_x\frac{\partial\beta}{\partial v}+\frac{\beta}{\kappa^2}\left(\frac{\partial^2H}{\partial u_x^2}-\frac{\partial^2H}{\partial v_x^2}\right)=0.
\end{equation}
Recalling from the third relation of \eqref{FGconstraint1} that $\beta^2-\alpha^2 $ is constant it follows from equations \eqref{firstpair} and \eqref{secondpair} that each of $\alpha$ and $\beta$ is constant and
\begin{equation}\label{Heq1}
  \frac{\partial^2H}{\partial u_x^2}-\frac{\partial^2H}{\partial v_x^2}=0, \quad \alpha\frac{\partial^2H}{\partial u_x^2}-\beta\frac{\partial^2 H}{\partial u_x \partial v_x}=0.
\end{equation}
Hence, $H$ is a quadratic function of $u_x, v_x$ of the form
\begin{equation}\label{Hexpression}
  H=\frac{\sigma}{2}\left(\beta u_x^2+2\alpha u_x v_x+\beta v_x^2\right)+\rho_u u_x +\rho_v v_x +\tau,
\end{equation}
where $\sigma,\rho_u,\rho_v,\tau$ are functions of $u,v$ only. Note also, $\sigma$ has the dimensions of velocity while $\rho_u,\ \rho_v$ have the same dimensions as the fields $u,v$, and $\tau$ has the dimensions of energy.
In addition, the third and fourth relations are now identical and given by:
\begin{equation}\label{fgeq}
 \frac{\partial f}{\partial v}- \frac{\partial g}{\partial u} -\frac{1}{\kappa^2}\left(\beta\frac{\partial^2H}{\partial v_x^2}-\alpha\frac{\partial^2 H}{\partial u_x \partial v_x}\right)\equiv \frac{\partial f}{\partial v}- \frac{\partial g}{\partial u} -\frac{\sigma}{c^2}=0.
\end{equation}
Note, this constraint on $f$ and $g$ is invariant under the \lq gauge' transformation \eqref{fggauge}. Expressions for $f$ and $g$ will be chosen later in section \ref{Summary}.
The requirement that $\alpha$ and $\beta$ are constant also simplifies the expressions for the derivatives of $P_D$. Explicitly, they are now given by:
\begin{eqnarray}\label{PDuvdervsagain}
 \nonumber  &&  \frac{\partial P_D}{\partial u}=u_x \frac{\partial f}{\partial u}+v_x \frac{\partial f}{\partial v} +\frac{1}{\kappa^2}\left(\alpha\,\frac{\partial H}{\partial u_x}-\beta\,\frac{\partial H}{\partial v_x}\right),\\
 \nonumber  &&\frac{\partial P_D}{\partial v}=u_x \frac{\partial g}{\partial u}+v_x \frac{\partial g}{\partial v}+\frac{1}{\kappa^2}\left(\beta\,\frac{\partial H}{\partial u_x}-\alpha\,\frac{\partial H}{\partial v_x}\right),\\
\nonumber &&\frac{\partial P_D}{\partial u_x}\equiv F =\alpha u_x +\beta v_x +f,\\
&& \frac{\partial P_D}{\partial v_x}\equiv G=\beta u_x +\alpha v_x +g .
\end{eqnarray}
The remaining requirements needed for matching the mixed $(u,v)$ derivatives will be considered later  once more information is known about $H$.

\subsection{The terms without time derivatives}\label{H}

\p The remaining terms to be considered are the terms that do not contain any time derivatives and these must cancel among themselves. The constraint arising from these terms in eq\eqref{Ptwithansatz} is
\begin{equation}\label{notconstraint}
  -\frac{1}{2}u_x^2+\frac{1}{2}v_x^2-\frac{\lambda}{3}u_x^3 + \frac{\lambda}{3}v_x^3 +u_x\frac{\partial H}{\partial u} + v_x\frac{\partial H}{\partial v}+\frac{1}{2\kappa^2}\left[\left(\frac{\partial H}{\partial u_x}\right)^2-\left(\frac{\partial H}{\partial v_x}\right)^2\right]=0.
\end{equation}
Using the expression for $H$ provided by eq\eqref{Hexpression}, the terms cubic in space derivatives in \eqref{notconstraint} lead to:
\begin{eqnarray}\label{threedervsa}
 \phantom{mmmm}u_x^3\ :&&\quad \frac{\partial \sigma}{\partial u}=\frac{2\lambda}{3\beta},\phantom{mmmmm} v_x^3\ : \quad \frac{\partial \sigma}{\partial v}=-\frac{2\lambda}{3\beta},  \\ \label{threedervsb}
u_x^2 v_x\ :&& \alpha \frac{\partial \sigma}{\partial u}=-\frac{\beta}{2}\frac{\partial \sigma}{\partial v},\quad\quad v_x^2 u_x\ : \quad\alpha \frac{\partial \sigma}{\partial v}=-\frac{\beta}{2}\frac{\partial \sigma}{\partial u}.
\end{eqnarray}
At this stage it is necessary to distinguish two cases, $\lambda=0$, and $\lambda\ne 0$.

\p If $\lambda=0$, which means the nonlinear term in the Boussinesq equation is temporarily disregarded, eqs\eqref{threedervsa} imply that $\sigma=\sigma_0$, a constant, and there is no additional constraint on $\alpha$, $\beta$ or $\sigma$ from eqs\eqref{threedervsb}. This case will be examined further in section \ref{linear}.

\p On the other hand, when $\lambda\ne 0$, the equations \eqref{threedervsa} and \eqref{threedervsb} imply the extra constraint $\beta=2\alpha$. Thus, for $\lambda\ne0$ and using the last relation of eq\eqref{FandG1}, $\alpha$, $\beta$ and $\sigma$ are given by:
\begin{equation}\label{alphabetasigma}
  \alpha^2=\frac{\kappa^2}{3c^2},\quad \beta^2=\frac{4\kappa^2}{3c^2}, \quad \sigma=\sigma_0+\mu\,(u-v), \ \ \mu=\frac{2\lambda}{3\beta}.
\end{equation}
The three terms in \eqref{notconstraint} quadratic in space derivatives lead to
\begin{equation}\label{twodervs}
u_x^2\ :\ \  \frac{\partial \rho_u}{\partial u}=\frac{1}{2}\left(1-\frac{\sigma^2}{c^2}\right),\ \  v_x^2\ :
  \ \  \frac{\partial \rho_v}{\partial v}=-\frac{1}{2}\left(1-\frac{\sigma^2}{c^2}\right),\quad u_xv_x\ : \ \  \frac{\partial \rho_v}{\partial u}=-\frac{\partial \rho_u}{\partial v},
\end{equation}
while the two terms linear in space derivatives in \eqref{notconstraint} are
\begin{equation}\label{onederv}
  u_x\ :\quad \frac{\partial\tau}{\partial u}+\frac{\sigma}{\kappa^2}\left(\beta\,\rho_u-\alpha\,\rho_v\right)=0,\quad v_x\ : \quad \frac{\partial\tau}{\partial v}+\frac{\sigma}{\kappa^2}\left(\alpha\,\rho_u-\beta\,\rho_v\right)=0.
\end{equation}
Finally, the terms independent of $u_x$ and $ v_x$ in \eqref{notconstraint} must also cancel, implying
\begin{equation}\label{noderv}
  \rho_u^2=\rho_v^2.
\end{equation}

\p Then, it follows from \eqref{twodervs}, \eqref{threedervsa} and \eqref{threedervsb} that
$$\frac{\partial^2 \rho_u}{\partial u\partial v}=-\frac{\sigma}{c^2}\frac{\partial\sigma}{\partial v}=\frac{\mu\sigma}{ c^2}, \quad \frac{\partial^2 \rho_v}{\partial v\partial u}=\frac{\sigma}{c^2}\frac{\partial\sigma}{\partial u}=\frac{\mu\sigma}{ c^2},$$
which is not compatible with the choice $\rho_u=-\rho_v$. Hence,  \eqref{noderv} implies $$\rho_u=\rho_v\equiv \rho,$$ and the last relation of \eqref{twodervs} implies $\rho$ is a function of the combination $u-v$. Then, the first two relations of \eqref{twodervs} are identical and on integration imply
\begin{equation}\label{rhoexpression}
  \rho=
  \rho_0+\frac{\sigma}{2\mu}\left(1-\frac{\sigma^2}{3c^2}\right).
\end{equation}

\p At this stage eqs\eqref{onederv} become
$$\frac{\partial\tau}{\partial u}=-\frac{\beta\sigma\rho}{2\kappa^2}=-\frac{\partial\tau}{\partial v},$$
implying that $\tau$ is also a function of $u-v$. Integrating these expressions gives
\begin{equation}\label{tauexpression}
\tau=\tau_0-\frac{\beta\sigma^2}{4\kappa^2\mu^2}\left(\mu\rho_0+\frac{\sigma}{3}\left(1-\frac{\sigma^2}{5 c^2}\right)\right).
  \end{equation}
  Again, this is a compact expression but the constants would need to be redefined if taking the limit $\mu\rightarrow 0$. In any case, adding a constant to $H$ simply adds a constant to ${\cal D}$, which can be ignored. Note also, the expression \eqref{tauexpression} is a fifth order polynomial in the discontinuity $u-v$, which is very reminiscent of the explicit B\"acklund transformation for the Boussinesq equation presented by Huang \cite{h1982}. This should not be a surprise since B\"acklund transformations are known to be intimately related to this type of defect (see \cite{bcz2005} for details of the sine-Gordon example). It is also reminiscent of the structure of the sewing relations describing a similar defect within the KdV model \cite{cz2005}, which also contain a fifth order polynomial in $u-v$.

 \p  At this stage, since $H$ is determined, the $(u,v)$ compatibility relation mentioned at the end of  section (3.2), but not analysed there, is seen to follow automatically on using the explicit form of $H$ together with the constraint on $f$ and $g$ given by eq\eqref{fgeq}. Note, the constant terms in $\tau$ can be discarded in the defect contribution to the action since they play no role.

 \subsection{Summary}\label{Summary}

\p At this stage, the various components of the defect contribution can be brought together and summarised (for $\lambda\ne 0$):
\begin{eqnarray}\label{Dsummary}
\nonumber &&{\cal D}=u_t\, F +v_t G +H, \ \ F=\alpha u_x +\beta v_x +f,\ \ G=\beta u_x +\alpha v_x +g, \ \beta^2=\frac{4\kappa^2}{3c^2},\ \alpha^2=\frac{\kappa^2}{3c^2}\,;\\
\nonumber&&H=\frac{\sigma}{2}\left(\beta u_x^2 +2\alpha u_x v_x +\beta v_x^2\right)+\rho (u_x+v_x)+\tau,\ \ \sigma=\sigma_0+\mu(u-v),\ \  \mu=\frac{2\lambda}{3\beta},\\
\nonumber&&\phantom{mmm} \rho=\rho_0+\frac{\sigma}{2\mu}\left(1-\frac{\sigma^2}{3c^2}\right), \ \ \tau=\tau_0-\frac{\beta\sigma^2}{4\kappa^2\mu^2}\left(\mu\rho_0+
\frac{\sigma}{3}\left(1-\frac{\sigma^2}{5c^2}\right)\right),\ \\
&&\phantom{mmmm}\ \frac{\partial f}{\partial v}-\frac{\partial g}{\partial u}=\frac{\sigma}{c^2};\ \ \hbox{for\ example:}\ \  f=g=-\frac{\sigma^2}{2\mu c^2} \ \ \hbox{(but not uniquely)}. \end{eqnarray}
Using these expressions in eqs\eqref{uvconditions} provides the set of explicit sewing conditions that must be satisfied by the fields $u,v$ and their derivatives at the defect location $x=x_0$. These are:
\begin{eqnarray}\label{uvconditions-summary}
\nonumber &\kappa^2\, u_{xx}&=\phantom{- l}\alpha u_t +\beta v_t +\sigma(\beta u_x +\alpha v_x)+\rho,\\
\nonumber &\kappa^2\, v_{xx}&=-\left(\beta u_t +\alpha v_t +\sigma(\alpha u_x +\beta v_x)+\rho\right),\\
\nonumber &\kappa^2 u_{xxx}&=\phantom{-} u_x(1+\lambda u_x)+\frac{\sigma}{c^2}\, v_t +\alpha u_{xt}+\beta v_{xt}-\frac{\mu}{2}(\beta u_x^2 +2 \alpha u_x v_x +\beta v_x^2)\\ && \nonumber \phantom{mmmmm}-\frac{1}{2}\left(1-\frac{\sigma^2}{c^2}\right)(u_x +v_x)+\frac{b\sigma}{2\kappa^2}\left(\rho_0+\frac{\sigma}{2\mu}-\frac{\sigma^3}{6\mu c^2}\right),\\
\nonumber &\kappa^2 v_{xxx}&=\phantom{-} v_x(1+\lambda v_x)+\frac{\sigma}{c^2}\, u_t -\beta u_{xt}-\alpha v_{xt}-\frac{\mu}{2}(\beta u_x^2 +2 \alpha u_x v_x +\beta v_x^2)\\ && \phantom{mmmmm}-\frac{1}{2}\left(1-\frac{\sigma^2}{c^2}\right)(u_x +v_x)+\frac{b\sigma}{2\kappa^2}\left(\rho_0+\frac{\sigma}{2\mu}-\frac{\sigma^3}{6\mu c^2}\right).
\end{eqnarray}

\subsection{An expression for the contribution to momentum from the defect}

\p Since the eqs\eqref{PDuvdervsagain} are consistent they can be integrated to determine an expression for $P_D$ via \eqref{Ptrearranged}.  Thus,
\begin{equation}\label{PDexpression}
  P_D=\frac{1}{2}\left(\alpha u_x^2 +2\beta v_x u_x +\alpha v_x^2\right) +f u_x +g v_x +h, \quad \frac{\partial h}{\partial u}=-\frac{\partial h}{\partial v}=-\frac{(\beta-\alpha) \rho}{\kappa^2}.
\end{equation}
Given the expression \eqref{rhoexpression} for $\rho$, the function $h(u,v)$ is given explicitly by
\begin{equation}\label{hexpression}
  h=h_0-\frac{(\beta-\alpha)\sigma}{\kappa^2\mu^2}\left(\mu\rho_0+\frac{\sigma}{4}\left(1-\frac{\sigma^2}{6 c^2}\right)\right),
\end{equation}
though $h_0$ and other constant terms in this expression are inessential. On the other hand, $f$ and $g$ are not uniquely determined for the reasons mentioned previously. A possible choice would be to take
\begin{equation}\label{fgpossibility}
  f=g=-\frac{\sigma^2}{2\mu c^2}.
\end{equation}
However, constant terms in this expression would play no role in the defect action. The undetermined parameters that do play a role in the defect action (for $\lambda\ne 0$) are $\sigma_0$ and $\rho_0$.

\p Once ${\cal D}$ is determined, the expressions for the conserved energy \eqref{Etotal} and conserved momentum are given by:
\begin{eqnarray}\label{conserevdEandP}
\nonumber && E_{\rm tot}=E+u_t\frac{\partial{\cal D}}{\partial u_t}+v_t\frac{\partial{\cal D}}{\partial v_t}-{\cal D}\equiv E-H\\ &&\phantom{mm}=E_u+E_v -\left[\frac{\sigma}{2}\left(\beta u_x^2 +2\alpha u_x v_x +\beta v_x^2\right)+\rho (u_x+v_x)+\tau\right]_{x=x_0},\\
\nonumber && P_{\rm tot}=P-u_x\frac{\partial{\cal D}}{\partial u_t}-v_x\frac{\partial{\cal D}}{\partial v_t}+P_{\rm D}\\ && \phantom{mm}= P_u+P_v-\left[\frac{1}2\left(\alpha u_x^2+2\beta u_x v_x +\alpha v_x^2\right)-h\right]_{x=x_0}.
\end{eqnarray}
Alternatively, the explicit contributions $E_{\cal D}$ and $P_{\cal D}$ are seen to be
\begin{eqnarray}
&&\label{ED} E_{\cal D}= -\left[\frac{\sigma}{2}\left(\beta u_x^2 +2\alpha u_x v_x +\beta v_x^2\right)+\rho (u_x+v_x)+\tau\right]_{x=x_0},\\
&&\label{PD} P_{\cal D}=-\left[\frac{1}2\left(\alpha u_x^2+2\beta u_x v_x +\alpha v_x^2\right)-h\right]_{x=x_0}.
\end{eqnarray}

\section{The linear case with $\lambda=0$}\label{linear}

\p  As noted in section \ref{H}, when  $\lambda =0$ it follows that $\sigma=\sigma_0$, a constant, and $\alpha$, $\beta$ only satisfy the quadratic constraint given in \eqref{FandG1}.\\

\p {\bf (i)}  The simplest choice would be to take $\sigma_0=c$, which requires $\rho_u=\rho_v\equiv\rho$ to be constant via \eqref{twodervs}. Taking $\rho=0$, $\tau=0$, and choosing
$$f=\frac{v}{2c},\ \ g=-\frac{u}{2c}$$
as a solution to \eqref{fgeq}, gives the following expression for the defect contribution:
\begin{equation}\label{Dlambdazeroex}
  {\cal D}_0=\frac{1}{2c}(u_t v- v_t u) + u_t(\alpha u_x +\beta v_x)+v_t(\beta u_x +\alpha v_x)+\frac{c}{2}\left(\beta u_x^2 +2\alpha u_x v_x +\beta v_x^2\right).
\end{equation}
In this case, the conserved energy and momentum, including the defect contributions, are given by
\begin{equation}\label{}
  E_0=E-\frac{c}{2}\left(\beta u_x^2 +2\alpha u_x v_x +\beta v_x^2\right),\ \ P_0=P-\frac{1}2\left(\alpha u_x^2+2\beta u_x v_x +\alpha v_x^2\right),\ \ \beta^2-\alpha^2 = \frac{\kappa^2}{c^2}.
\end{equation} Thus, for this particular example the expressions for $E_0$ and $P_0$ are nicely symmetrical.\\

\p {\bf (ii)} However, if $\sigma_0\ne c$ it would be natural to take
$$f=\frac{\sigma_0v}{2c^2}, \ \ g=-\frac{\sigma_0 u}{2c^2},$$ as solutions to \eqref{fgeq}, and the other pieces $\rho_u, \rho_v$ and $\tau$ would not be constants. Rather,  eq\eqref{noderv} allows two different possibilities: either (a) $\rho_u= \rho_v\equiv \rho$, or (b) $\rho_u=-\rho_v\equiv\rho$. In option (a) $\rho_u$ and $\tau$ are functions of $u-v$, while in option (b) $\rho_u$ and $\tau$ are functions of $u+v$.  Thus, for option (a), for example, the expressions are:
\begin{equation}\label{}
  \rho=\rho_0+\frac{1}{2}\left(1-\frac{\sigma_0^2}{c^2}\right)\, (u-v),\ \ \tau=\frac{\sigma_0(\alpha-\beta)}{\kappa^2}\left[\rho_0(u-v)+\frac{1}{4}\left(1-\frac{\sigma_0^2}{c^2}\right)(u-v)^2\right].
\end{equation}
Also, using the last pair of eqs\eqref{PDexpression}, $h(u,v)$ is given by
\begin{equation}\label{}
  h=\frac{(\alpha-\beta)}{\kappa^2}\left[\rho_0(u-v)+\frac{1}{4}\left(1-\frac{\sigma_0^2}{c^2}\right)(u-v)^2\right].
\end{equation}
In this case, the defect contribution to the action is more elaborate. As before, $\rho_0$ and $\sigma_0$ are constants playing a role but $h_0$ and $\tau_0$ are irrelevant and omitted.

\p The next step is to analyse how the defect affects a linear wave (in the cases for which $\lambda=0$), or a soliton (in the cases for which $\lambda\ne 0$). The latter is more interesting but the linear wave will be considered first. From previous experience, it might expected that the defect is purely transmitting in all cases.

\subsection{A linear wave encountering a defect}

\p The simplest case to consider has $\lambda=0$, $\rho_0=0$, places the defect at $x=0$, and uses the ansatz
\begin{equation}\label{linearwave}
u=\left(e^{ikx}+Re^{-ikx}\right)\, e^{-i\omega t},\ \ \ v=T e^{ikx}\, e^{-i\omega t},\ \omega>0,\ k>0,\  \omega^2=c^2 k^2(1+\kappa^2 k^2),
\end{equation}
where the reflection coefficient $R$, and transmission coefficient $T$, are to be determined by the sewing condtions \eqref{uvconditions}. In this case, the expression for ${\cal D}$ is assembled using the data in part (ii) above. However, $\rho_0$ has been taken to be zero because otherwise it would correspond to a constant shift in one or both of the expressions for $u,v$ given above. The sewing conditions for option (a), for example, lead to four equations for $R$ and $T$. Nevertheless, the over-determined system for $R$ and $T$ has the unique solution
\begin{equation}\label{linearRT}
  R=0,\ \ T=-\,\frac{(\alpha+\beta) c^2 k^2 +i\sigma_0 k+i\omega}{(\alpha+\beta) c^2 k^2 -i\sigma_0 k-i\omega},\ \ T\bar{T}=1.
\end{equation}
This contains two tunable parameters, $\sigma_0$ and $\alpha$, on recalling that
$$\beta^2=\alpha^2+\frac{\kappa^2}{c^2}\,.$$
Clearly, the defect within the linearised Boussinesq equation is purely transmitting, as expected from previous experience. Choosing $\sigma_0=c,\ \beta=\kappa/c,\ \alpha=0$ leads to a particularly simple expression for a transmission factor, namely $$T_0=-\,\frac{ c\kappa k^2 +ic k+i\omega}{ c \kappa k^2 -i c k-i\omega}=\frac{ck}{\omega}\,(1-i\kappa k) \equiv -i\tanh\left(\frac{\theta}{2}+\frac{i\pi}{4}\right) ,$$
where the last equality has made use of a \lq rapidity' variable, setting $k=(1/\kappa)\sinh\theta,\ \omega=(c/2\kappa)\sinh 2\theta$,  to represent solutions to the dispersion relation  in  eq\eqref{linearwave}.

\section{Solitons encountering a defect}

\subsection{A single soliton} An expression for a single soliton solution (in terms of $u$ satisfying \eqref{BZueq}) of the Boussinesq equation \eqref{BZweq} is given in \eqref{Bsoliton}. For this section, it is useful to reorganise the expression as follows:
\begin{equation}\label{Eusoliton}
w=u_x,\ \ u=u_0+\frac{3\kappa\gamma}{\lambda}\,\left(\frac{1-E}{1+E}\right),\ \ E=e^{\gamma(x-\nu t)/\kappa}, \ \gamma=\sqrt{1-\nu^2/c^2},\ \ |\nu|<c.
\end{equation}
Also, because $u_0$ is arbitrary, adjusting it leads to an alternative, slightly simpler but equivalent, expression given by:
 \begin{equation}\label{Eusolitona}
 u=u^\prime_0-\frac{6\kappa\gamma}{\lambda}\,\left(\frac{E}{1+E}\right),\ \ E=e^{\gamma(x-\nu t)/\kappa}, \ \gamma=\sqrt{1-\nu^2/c^2},\ \ |\nu|<c.
\end{equation}
Since the sewing conditions at the defect location must hold for all time, it is sensible to take the expression \eqref{Eusolitona} in the region $x<x_0$, and to suppose the soliton solution (in terms of $v$) in the region $x>x_0$ is given by
\begin{equation}\label{Evsoliton}
 v=v_0+\frac{3\kappa\gamma}{\lambda}\,\left(\frac{1-zE}{1+zE}\right), \ \hbox{or},\ v=v^\prime_0-\frac{6\kappa\gamma}{\lambda}\,\left(\frac{zE}{1+zE}\right),
 \end{equation}
where the parameter $z$ is to be determined. The shifts $u_0$ or $u^\prime_0$ and $v_0$ or $v^\prime_0$ can be removed entirely provided  $\rho_0$ is given by
\begin{equation}\label{rho0expression}\rho_0=-\frac{\sigma_0}{2\mu}\left(1-\frac{\sigma_0^2}{3c^2}\right).\end{equation}
As in the linear case, the quantity $z$ is over-determined and it is necessary to check that there is a consistent solution to the four sewing conditions. This is straightforward, though quite lengthy, and the necessary algebra has been handled using Maple.

\p To obtain a simple expression for $z$, it is useful to express the speed of the soliton in terms of rapidity by setting $\nu =c\tanh\theta$,  and define the essential parameter $\sigma_0$ (which is not necessarily constrained in magnitude to be less than $c$) as a multiple of $c$, namely $\sigma_0=s c$. Then, $z$ is given by
\begin{equation}\label{zexpression}
 z=\frac{\sinh\theta + s \cosh\theta\mp\sqrt{3}}{\sinh\theta +s \cosh\theta \pm\sqrt{3}}, \ \ \alpha=\pm\,\frac{\kappa}{\sqrt{3}\, c}.
\end{equation}
The two possibilities arise because of the relations given in \eqref{alphabetasigma}, which imply two possible choices for $\alpha$.
Provided $z$ is positive, both $u$ and $v$ are non-singular solutions in their respective domains. Note, the expression for $z$ is independent of the location of the defect.

\begin{figure}[H]
  \centering
  \includegraphics[scale=0.55]{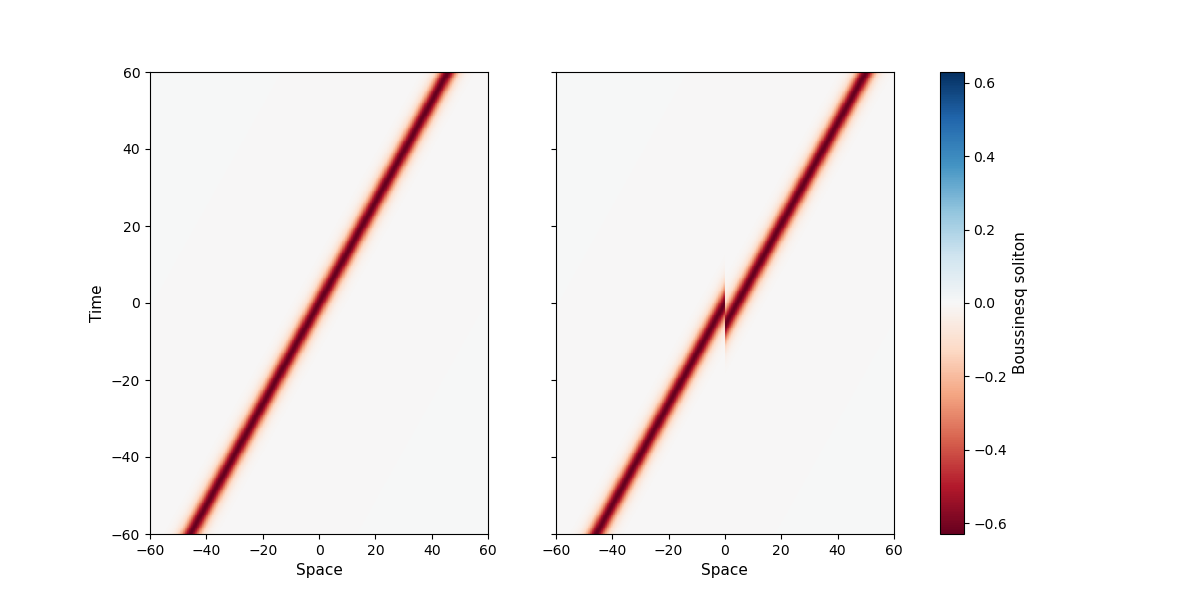}
  \caption{\it Soliton and a defect ($\theta=1.0, s=0.5$), the right hand colormap  shows the delay caused by the defect. Note: The left hand plot represents $u_x$ in the absence of a defect while the right hand plot displays $u_x$ for $ x<0 $ and $v_x$ for $x>0$ with a defect at $x=0$.}\label{solitoncolormap}
\end{figure}

\begin{figure}[H]
  \centering
  \includegraphics[scale=0.55]{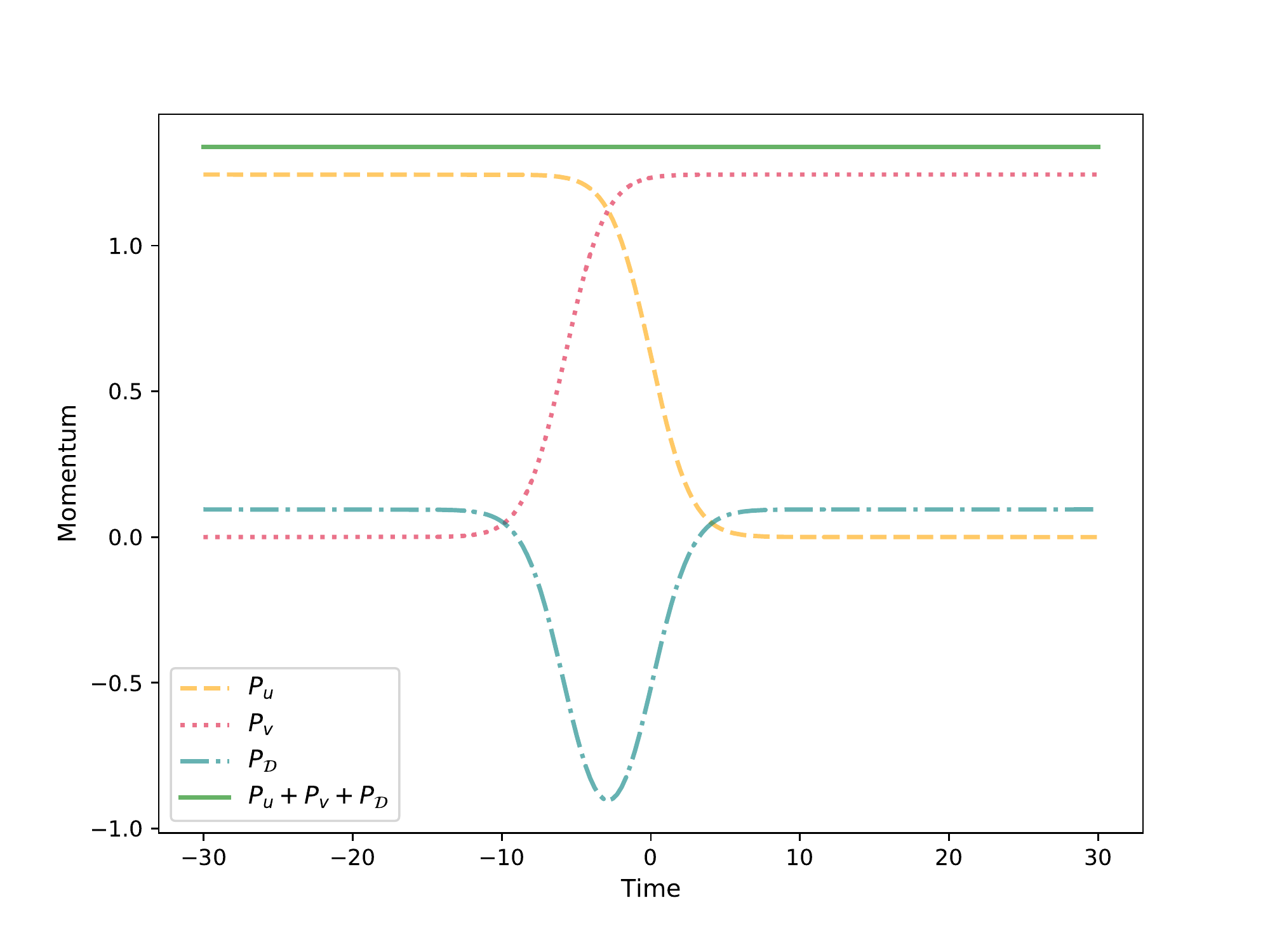}
  \caption{\it Momentum conservation for a single soliton with a defect ($\theta=1.0, s=0.5$).}\label{solitonmomentum}
\end{figure}

\p A typical situation is illustrated in Figs(\ref{solitoncolormap},\ref{solitonmomentum}) where a soliton meets a defect using the parameter choices $c=\lambda=\kappa=1, \ \theta =1.0,\ s=0.5$, the latter implying $z\approx 0.05836$ using \eqref{zexpression}.

\p In Fig(\ref{solitoncolormap}), the colormap on the left represents a soliton without the defect while the colormap on the right represents the effect of the defect. Note, in this type of picture, the defect position has been chosen to be $x_0=0$, $u_x$ is plotted on the left hand diagram and then $u_x,\ x<0,\ v_x, \ x>0$ are plotted on the right hand diagram. In other words, on the colormaps, it is the solutions to \eqref{BZweq} that are illustrated because the spatial extents of the solitons are more clearly evident.

\p In the second diagram, Fig(\ref{solitonmomentum}), the contributions to $P_u, P_v$ and $P_{\cal D}$, as defined in eq\eqref{PD}, are plotted as functions of time. It is clear from the plots that the soliton momentum is transferred across the defect and no part of it is stored there after the soliton has passed; the total momentum is conserved, as expected.

\p On the other hand, if the rapidity variable $\theta$ is chosen to ensure the numerator or denominator vanishes in the expression for $z$ given by \eqref{zexpression}, then the solution $v$ for $x>0$ is constant, meaning the soliton will have disappeared with its energy and momentum absorbed by the defect. However, this is not always possible. If $|s|>2$, the soliton rapidity cannot be chosen in this way.

\p Finally, if $z<0$ the solution $v$ develops a singularity.

\subsection{Two solitons}
With the conventions adopted here, a double soliton  solution \cite{h1973} is given by
\begin{equation}\label{twosoliton}
u=-\frac{6\kappa}{\lambda}\ \frac{\gamma_1 E_1+\gamma_2 E_2 +(\gamma_1+\gamma_2)\, S_{12}\, E_1 E_2}{1+ E_1+ E_2 + S_{12}\, E_1 E_2},
\end{equation}
where
\begin{equation}\label{S12}
\nu_1=c\tanh\theta_1,\ \nu_2=c\tanh \theta_2,\ \hbox{and}\ S_{12}=\frac{\sinh^2(\theta_1-\theta_2)-3 (\cosh \theta_1 -\cosh\theta_2)^2}{\sinh^2(\theta_1-\theta_2)-3 (\cosh \theta_1 +\cosh\theta_2)^2}\, .
\end{equation}
The latter expression for $S_{12}$ factors into
\begin{equation}\label{S12factor}
S_{12}=\left(\frac{3\cosh(\theta_1+\theta_2)-\cosh(\theta_1-\theta_2)-4}{3\cosh(\theta_1+
\theta_2)-\cosh(\theta_1-\theta_2)+4}\right)\, \tanh^2\left(\frac{\theta_1-\theta_2}{2}\right),
\end{equation}
and  the first factor in \eqref{S12factor} can be rewritten:
\begin{equation}
\left(\frac{\cosh\left(\frac{\theta_1-\theta_2}{2}\right)-\sqrt{3}\sinh\left(\frac{\theta_1+\theta_2}{2}\right)}
{\sinh\left(\frac{\theta_1-\theta_2}{2}\right)-\sqrt{3}\cosh\left(\frac{\theta_1+\theta_2}{2}\right)}\right) \, \left(\frac{\cosh\left(\frac{\theta_1-\theta_2}{2}\right)+\sqrt{3}\sinh\left(\frac{\theta_1+\theta_2}{2}\right)}
{\sinh\left(\frac{\theta_1-\theta_2}{2}\right)+\sqrt{3}\cosh\left(\frac{\theta_1+\theta_2}{2}\right)}\right).
\end{equation}
However, none of the factors of $S_{12}$ seem closely related to the expression for $z$ given in eq\eqref{zexpression}.

\p For comparison, consider the sine-Gordon example  \cite{bcz2005}. There, the scattering of a soliton with a defect can be calculated similarly and the corresponding quantity $z$ bears a close relationship with the delay factor $A_{12}(\theta_1-\theta_2)$ describing the scattering of two solitons. In fact, performing the analogous calculations in the sine-Gordon model and using a similar notation, with an appropriately defined defect parameter $ \eta$, the result is found to be:
\begin{equation}\label{sGz} z=\coth\left(\frac{\eta-\theta}{2}\right),\ \ z^2=A_{12}(\eta-\theta).
\end{equation}
  The situation in the present context is evidently more complicated, which was to be expected. After all, the Boussinesq equation is not Lorentz invariant and there is no reason why delays such as \eqref{S12factor} should be functions solely of rapidity differences.

\p If a double soliton \eqref{twosoliton} encounters a defect from the region $x<x_0$ then the matching solution in the region $x>x_0$ has the form
\begin{equation}\label{twosolitonv}
v=-\frac{6\kappa}{\lambda}\ \frac{\gamma_1 z_1 E_1+\gamma_2 z_2 E_2 +(\gamma_1+\gamma_2)\, S_{12}\, z_1 z_2 E_1 E_2}{1+ z_1 E_1+ z_2 E_2 + S_{12}\, z_1 z_2 E_1 E_2},
\end{equation}
where each of $z_1$ and $z_2$ are given by the expression \eqref{zexpression} with $\theta$ replaced by $\theta_1$ and $\theta_2$, respectively. Again, this is straightforward to check using the Maple package and demonstrates that the soliton components behave independently when meeting a defect.

\subsection{Merging and splitting solitons} As mentioned in the Introduction, the Boussinesq equation also permits solutions that represent merging, or splitting, solitons \cite{rs2017}. This type of solution is not shared with the other integrable nonlinear wave equations where defects have been explored. In fact, when negotiating a defect these solutions display interesting behaviours not encountered before.

\p In the conventions of this article, an example of this type of solution is provided by setting
\begin{equation}\label{mergingsols}
u=-\frac{6\kappa}{\lambda}\, \frac{\omega_1 E_1+\omega_2 E_2+\omega_3 E_3}{ E_1+ E_2+ E_3 },\ \ E_a=e^{\omega_a(x-\nu_a t)/\kappa},\ \ \nu_a=\sqrt{3}\, w_a\, c,\ \  \ a=1,2,3,
\end{equation}
where
\begin{equation}\label{lambdas}
w_1=\frac{1}{2}\left(\frac{1}{\cosh\theta}+\frac{\tanh\theta}{\sqrt{3}}\right), \ \ w_2=\frac{1}{2}\left(-\frac{1}{\cosh\theta}+\frac{\tanh\theta}{\sqrt{3}}\right),\ \  w_3=-\frac{\tanh\theta}{\sqrt{3}}.
\end{equation}
 Note, the parameters $w_1,w_2, w_3$ have been chosen to satisfy explicitly the necessary conditions $$\omega_1+\omega_2+\omega_3=0,\ \ \omega_1^2+\omega_2^2+\omega_3^2=1/2,$$ but written in terms of a single variable $\theta$ to facilitate a comparison with the calculations for a single soliton. The variable $\theta$ plays the role of a rapidity for  $E_3$ the third component of the expression \eqref{mergingsols}.

\p This type of solution is also compatible with the defect (which, for convenience of calculation, is placed at $x_0=0$). To see this, the sewing conditions are used to perform the matching, assuming a solution for $u(x,t),\ x<0,$ given by \eqref{mergingsols}, and a solution for $v(x,t),\ x>0,$ of the type
\begin{equation}
v=-\frac{6\kappa}{\lambda}\, \frac{z_1\,\omega_1 E_1+z_2 \,\omega_2 E_2+z_3\, \omega_3 E_3}{z_1\, E_1+ z_2\, E_2+z_3\, E_3 },\ \ z_3\ne 0.
\end{equation}
Then, for the two possible choices of $\alpha$, either
\begin{equation}\label{z1z2alphaplus}
z_1=z_3\,\frac{s \cosh\theta-2\sinh\theta}{s \cosh\theta +\sinh\theta +\sqrt{3}},\ \ \ z_2=z_3\frac{s \cosh\theta-2\sinh\theta}{s \cosh\theta +\sinh\theta -\sqrt{3}}\,, \ \hbox{if}\  \alpha=\frac{\kappa}{\sqrt{3}\,c},
\end{equation}
or
\begin{equation}\label{z1z2alphaminus}
z_1=z_3\,\frac{s \cosh\theta +\sinh\theta +\sqrt{3}}{s \cosh\theta-2\sinh\theta},\ \ \ z_2=z_3\frac{s \cosh\theta +\sinh\theta -\sqrt{3}}{s \cosh\theta-2\sinh\theta}\,, \ \hbox{if}\ \alpha=-\frac{\kappa}{\sqrt{3}\,c}.
\end{equation}
It is interesting to note that in both cases the previously determined soliton transmission factor provided in \eqref{zexpression} is given by $z=z_1/z_2$. As before the Maple package is indispensable when checking the formulae for $z_1$ and $z_2$.

\p For most choices of the parameters $\theta$ and $s$, and provided $s\ne 2\tanh\theta$, there will be both reflection and transmission at the defect. To illustrate this, it is useful to plot separately the various contributions to the total momentum over time. As before, in the numerical calculations the Boussinesq parameters are chosen to be $c=\lambda=\kappa=1$. There are a number of cases to explore and a selection of possibilities are listed next.

\begin{figure}[H]
  \centering
  \includegraphics[scale=0.55]{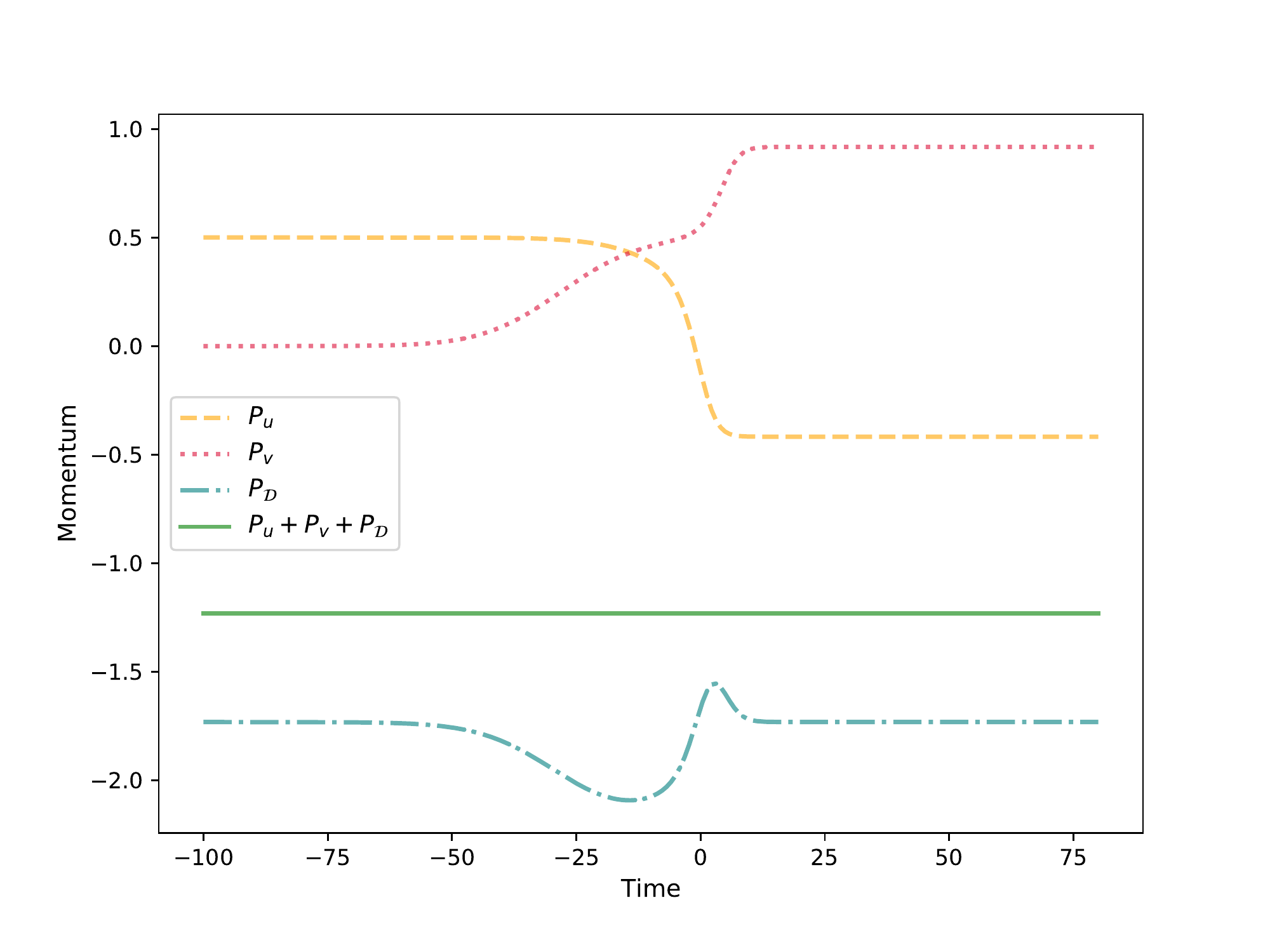}
  \caption{\it A splitting soliton interacting with a defect showing transmission and reflection illustrating momentum conservation ($\theta=-1.5,\ s=2.0$).}\label{mergingsolitontransmissionreflection}
\end{figure}

\p {\bf (i)}  Consider the case with $\alpha>0$ for which eq\eqref{z1z2alphaplus} is relevant. Then, Fig(\ref{mergingsolitontransmissionreflection}), with $\theta=-1.5$ and  $s=2$, illustrates a typical situation that involves transmission and reflection.  The plot clearly demonstrates  that the total conserved momentum (solid green line) is split between the ingoing soliton momentum $P_u$ and the defect before the defect is reached, and then, after the interaction with the defect, the momentum is split between the reflected soliton, the transmitted soliton, and the defect.

\p In this case, the momentum stored at the defect fluctuates during the interaction but returns to its pre-interaction value after the interaction.

\p The caption refers to a \lq splitting' soliton because even in the absence of a defect the intial soliton will split into two others, as was shown in \cite{rs2017}.

\p {\bf (ii)} However, if, in the case for which $|s|<2$,  $\theta$ is chosen  to satisfy $\tanh\theta = s/2$, then in the first option above, eq\eqref{z1z2alphaplus}, with $\alpha>0$, both $z_1$ and $z_2$ vanish, implying that $v$ is constant and nothing is transmitted into the region $x>0$.

\begin{figure}[H]
  \centering
  \includegraphics[scale=0.55]{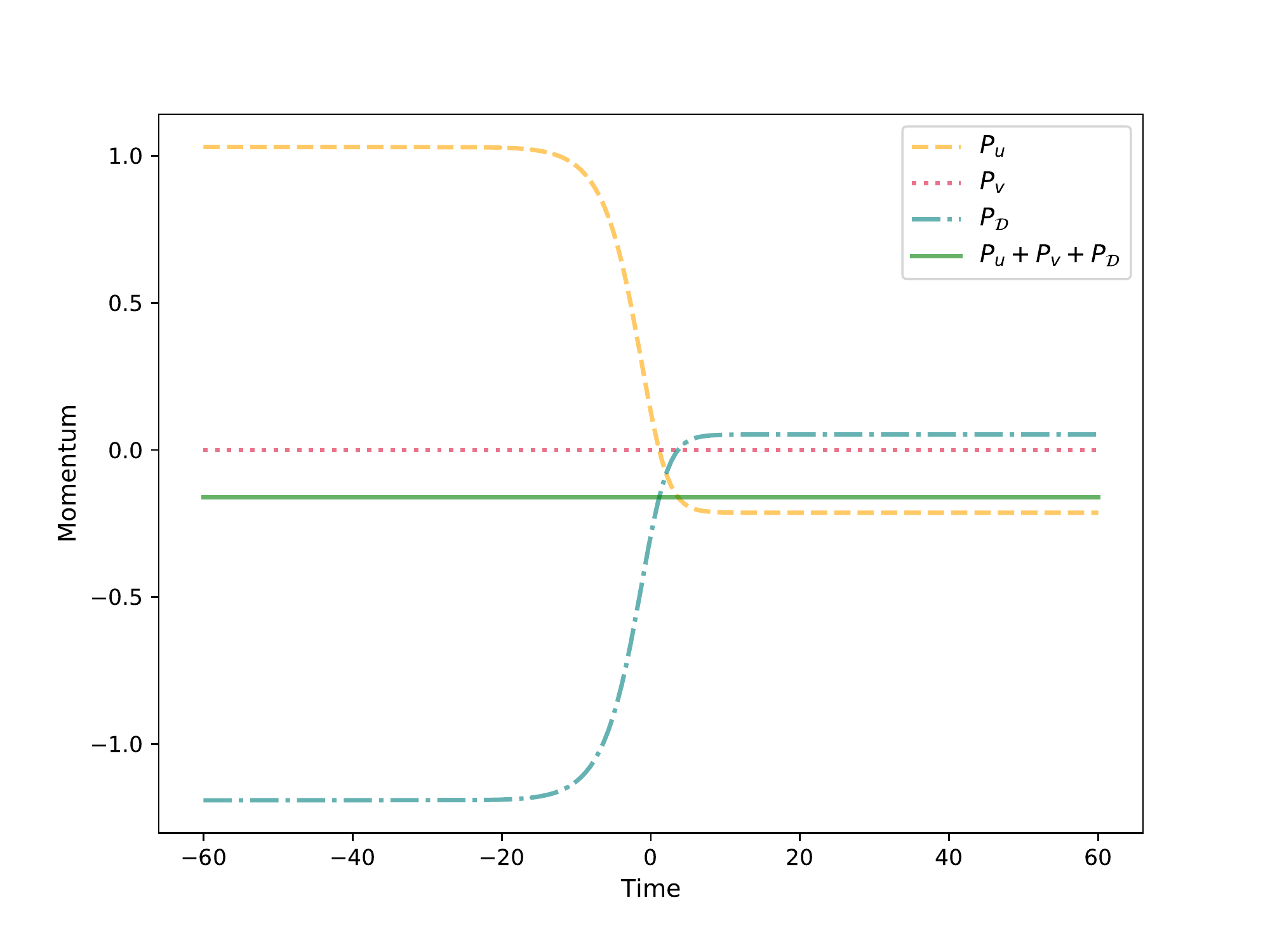}
  \caption{\it A merging soliton reflecting from a defect  ($z_1=z_2\equiv 0$  and $\theta=1.0$).}\label{mergingsolitonmomentum}
\end{figure}

\begin{figure}[H]
  \centering
  \includegraphics[scale=0.55]{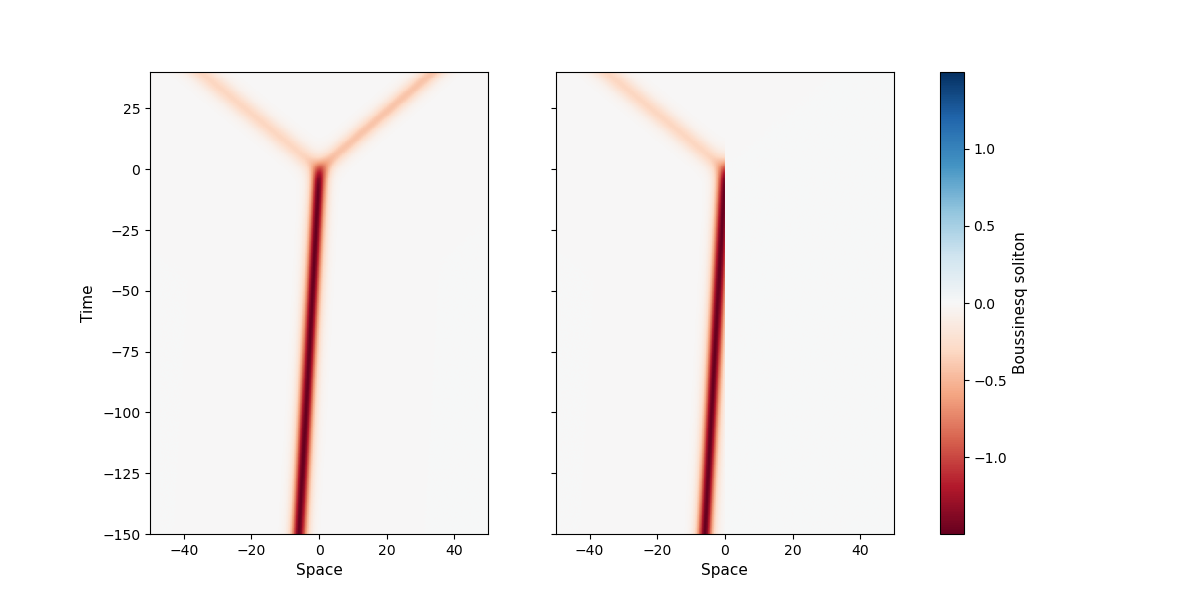}
  \caption{\it Colormap representing on the right hand diagram a reflecting soliton ($\theta=-1.4$).}\label{mergingsolitoncolormapwiththeta=-1.4}
\end{figure}

\begin{figure}[H]
  \centering
  \includegraphics[scale=0.55]{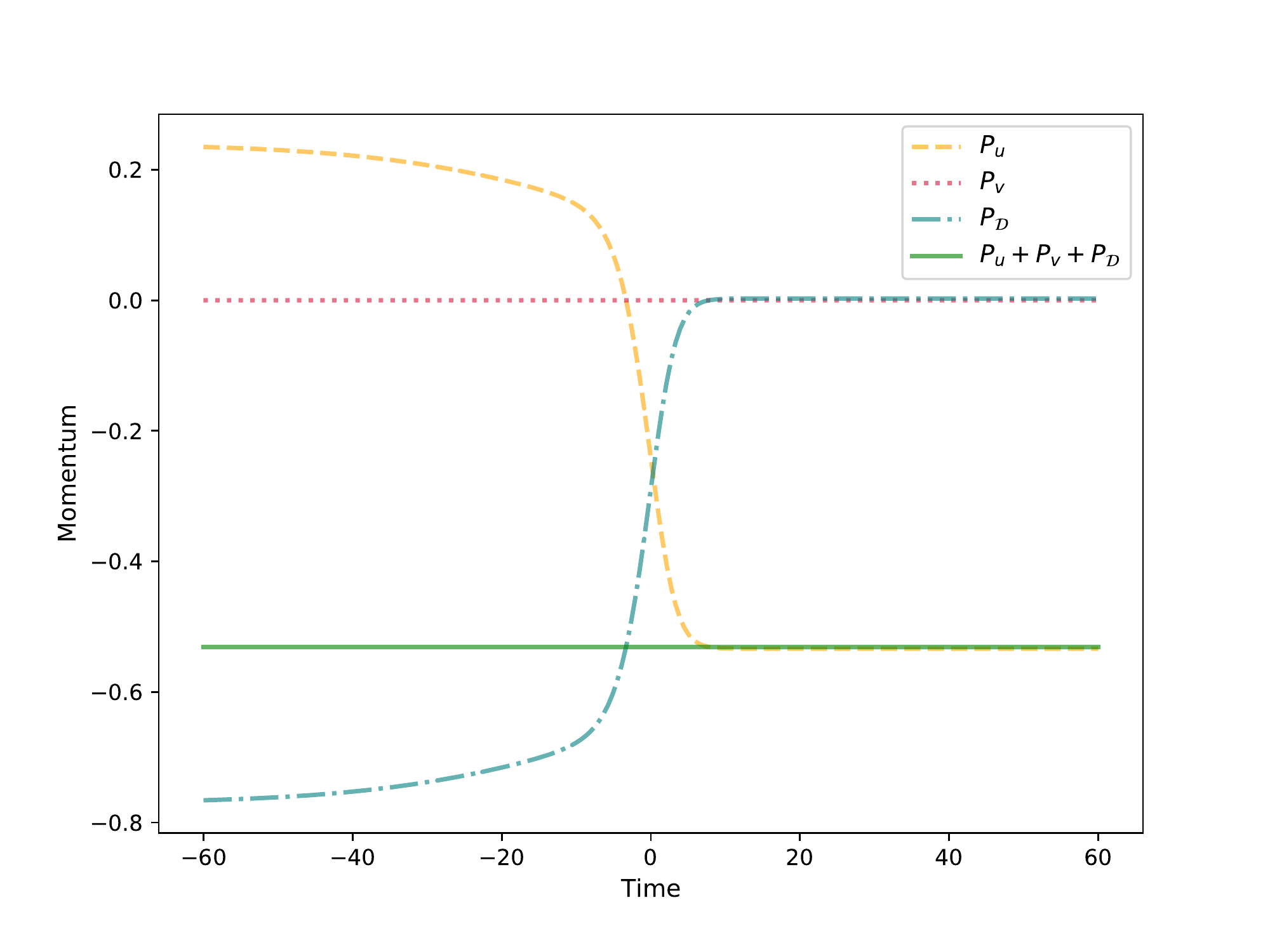}
  \caption{\it Conservation of momentum for a reflecting soliton ($z_1=z_2\equiv 0,\ \theta=-1.4$).}\label{mergingsolitonwiththeta=-1.4}
\end{figure}

\p  In this case, a soliton approaching the defect in the region $x<0$  might be reflected, with some of its energy and momentum stored by the defect, or captured with all its energy and momentum deposited on the defect. Illustrations of these possibilities are provided in Figs(\ref{mergingsolitonmomentum},\ref{mergingsolitoncolormapwiththeta=-1.4},\ref{mergingsolitonwiththeta=-1.4}).

\p In Fig(\ref{mergingsolitonmomentum}) with the choice $\theta=1.0$,  the reflected soliton shares its momentum with the defect and therefore reflects with a reduced momentum. In this case, there is a net change of momentum stored at the defect.

\p On the other hand, in Figs(\ref{mergingsolitoncolormapwiththeta=-1.4},\ref{mergingsolitonwiththeta=-1.4}), with the choice $\theta=-1.4$, the soliton appears to be almost perfectly reflected and there is no momentum stored in the defect after the interaction.

\p The colormap provided in the left hand diagram of Fig(\ref{mergingsolitoncolormapwiththeta=-1.4}) also illustrates the fact mentioned above that this type of soliton splits even without the defect. In this particular case, the defect is filtering out one of the emerging solitons following the split but allowing the other to carry momentum into the region $x<0$.

\begin{figure}[H]
  \centering
  \includegraphics[scale=0.55]{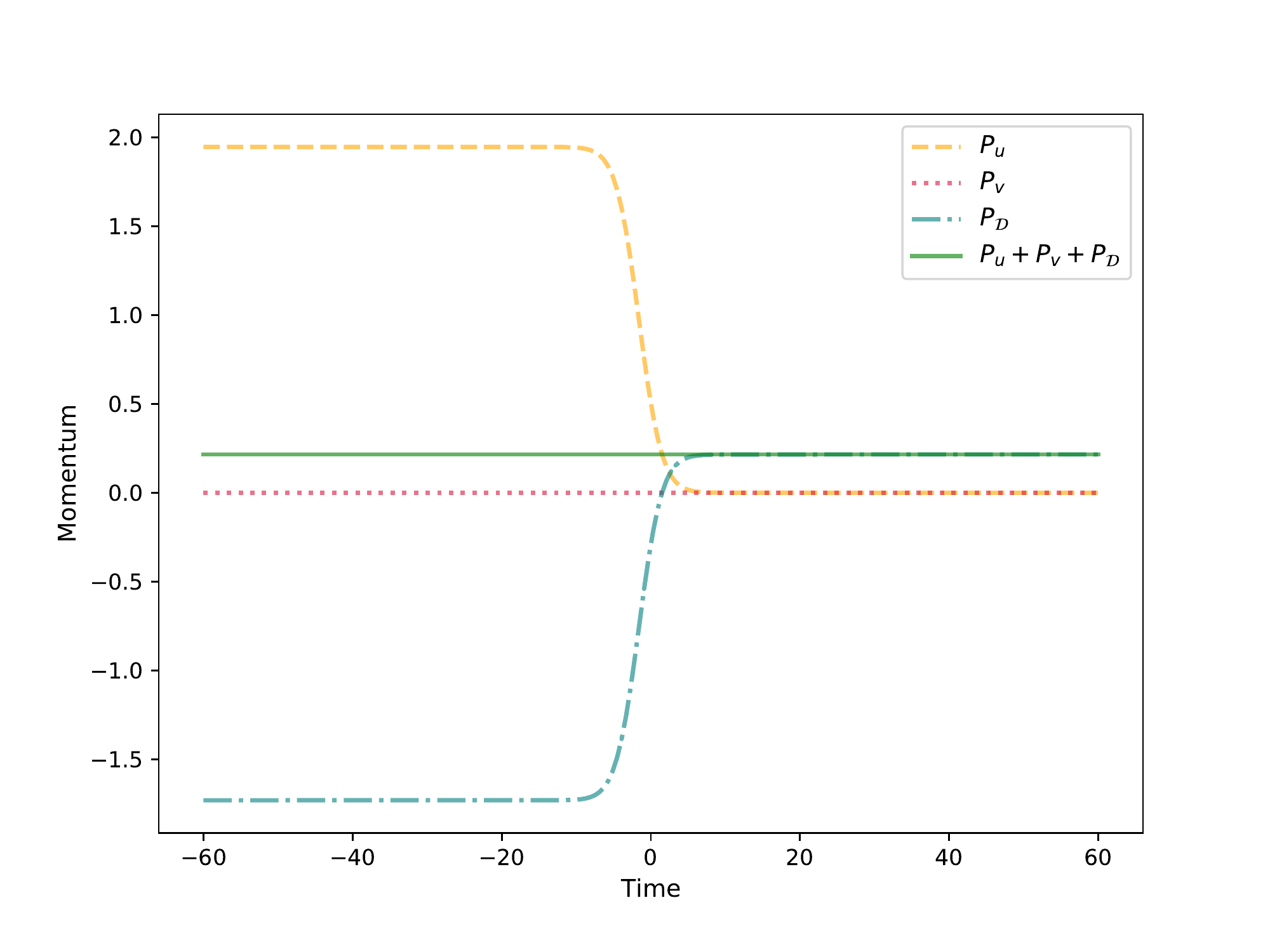}
  \caption{\it A \lq disappearing' soliton ($z_1=z_2\equiv 0,\ \theta=-5.0$).}\label{mergingsoliton-with-theta=-5}
\end{figure}

\p {\bf (iii)} Finally, still considering the case with $s=2\tanh\theta$, a soliton could be captured by the defect and this possibility occurs with $\theta$ in the range $-\infty<\theta\lesssim -3.0$ (the upper bound is not precise) for which $\omega_1$ and $\omega_2$ are both negative with $\omega_1 \approx \omega_2$. This is illustrated in Fig(\ref{mergingsoliton-with-theta=-5}), with $\theta=-5.0$. It is clear that almost all the momentum eventually resides in the defect with $P_u\rightarrow 0$ as $\theta$ decreases. Since the Boussinesq equation \eqref{BZueq} is invariant under time reversal, choosing $\theta=+5.0$ would lead to a situation where the defect is first set up with a discontinuity at early times and then evolves into a single soliton  carrying momentum away into the region $x<0$.

\p {\bf (iv)}  If the other option is adopted, which means $\alpha<0$ and $z_1$ and $z_2$ are defined instead by \eqref{z1z2alphaminus}, and $\theta$ is chosen  to satisfy $\tanh\theta = s/2$,  then both $z_1$ and $z_2$ diverge.

\p However, if the expression for $v(x,t)$ is first rearranged by multiplying the numerator and denominator by $s\cosh\theta-2\sinh\theta$ before setting $s=2\tanh\theta$, then the result is equivalent to setting  $z_3=0$ but keeping $z_1$ and $z_2$ finite. In that situation $v$  represents a soliton transmitted into the region beyond the defect with speed $\sqrt{3}(\omega_1+\omega_2)c\equiv c\tanh\theta$. Again there are several possibilities as the parameter $\theta$ is varied.

\begin{figure}[H]
  \centering
  \includegraphics[scale=0.55]{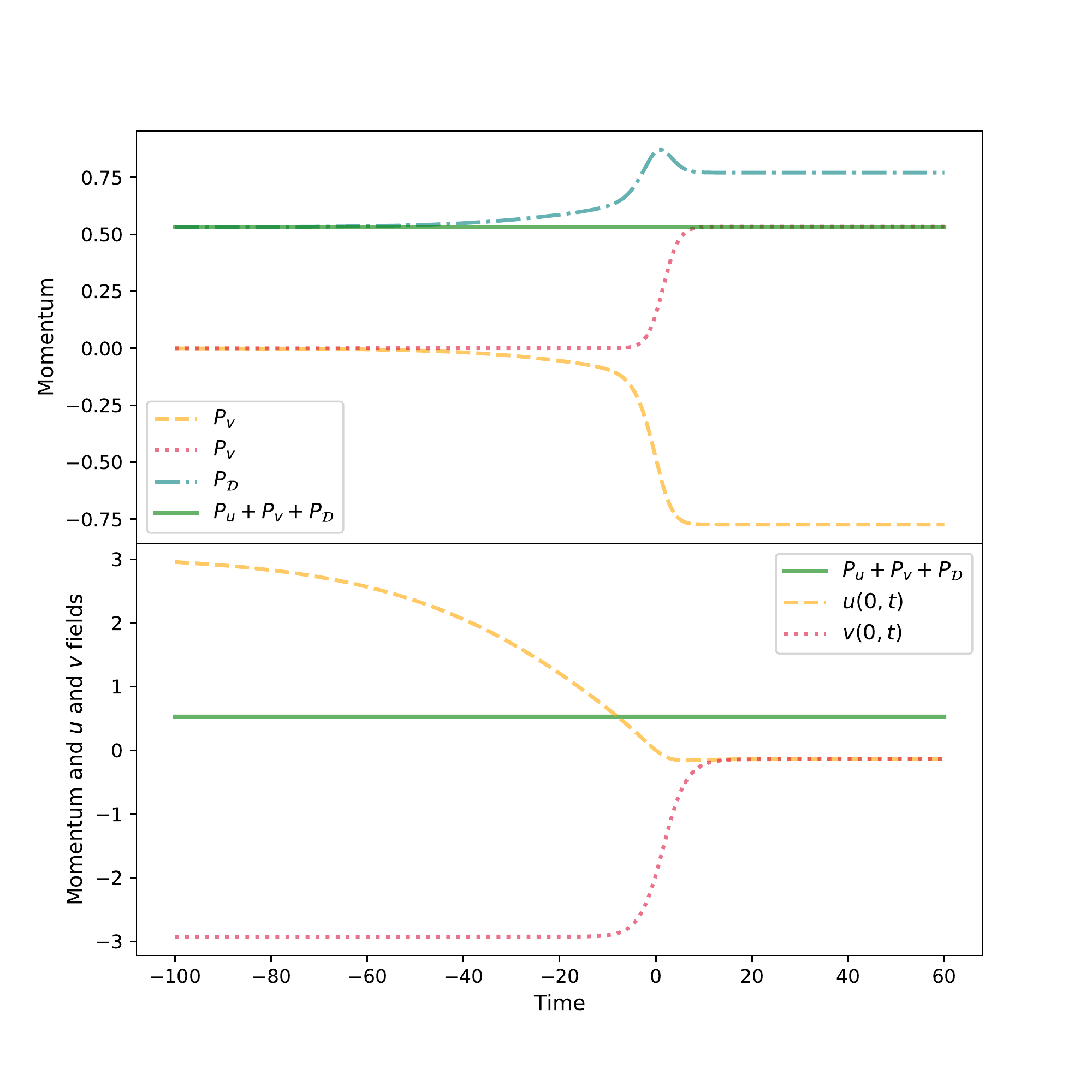}
  \caption{\it Momentum conservation with a \lq decaying' defect ($\theta=1.4$). The lower diagram indicates the initial discontinuity by plotting $u(0,t)$ and $v(0,t)$ and demonstrates that the discontinuity disappears as the two solitons emerge.}\label{decayingdefect-with-theta=1.4}
\end{figure}

\p In this parameter regime, one particularly intriguing situation deserves a special mention. It is possible to create an initial state with all the momentum stored in the defect (by arranging a discontinuity at early times), which then evolves into a pair of solitons. In order to conserve momentum one of the decay products is travelling in the direction of decreasing $x$, the other in the direction of increasing $x$. This possibility is illustrated in Fig(\ref{decayingdefect-with-theta=1.4}) where $u(0,t)$ and $v(0,t)$ are also indicated (though not to scale since they are dimensionally different to the momenta), to demonstrate how the initial discontinuity ultimately disappears.

\begin{figure}[H]
  \centering
  \includegraphics[scale=0.55]{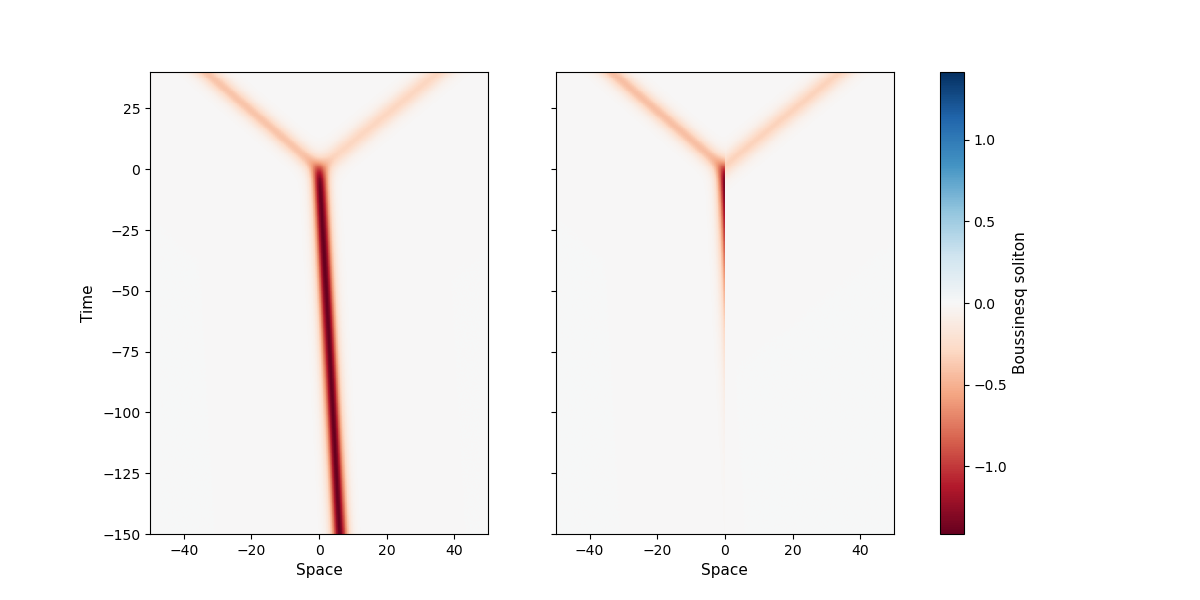}
 \caption{\it A \lq decaying' defect ($\theta=1.4$). The left hand diagram represents the soliton ($u_x$) without the defect, the right hand diagram represents the solitons $u_x$ for $x<0$ and $v_x$ for $x>0$ as they emerge from the defect travelling  in opposite directions.}\label{decayingdefectcolormap-with-theta=1.4}
\end{figure}

\p The colormap in Fig(\ref{decayingdefectcolormap-with-theta=1.4}) reveals how the defect manages this trick by converting the initial momentum of the splitting soliton arriving from the region $x>0$ into a discontinuity, which subsequently fades to zero as the two solitons emerge. However, the defect clearly gains momentum in this manoeuvre and this is stored in the final state. This  happens due to a dependence on the parameter $\sigma_0=sc$ in $h(0,t)$, eq\eqref{hexpression}, which is part of the defect contribution to the total momentum $P_{\cal D}$. Because of this, there is a contribution to the defect momentum that persists even when the discontinuity $u(0,t)-v(0,t)$ vanishes.

\p Taken together, this is an interesting collection of phenomena because in all other examples of nonlinear wave equations that support defects of the type considered here the defects have turned out to be purely transmitting. In those examples, a soliton might be captured for special choices of parameter, or (for example in the sine-Gordon model \cite{bcz2005}), a defect might convert a soliton to an anti-soliton while not altering its momentum. The phenomena described above in the Boussinesq context seem to provide the first examples of an integrable defect exhibiting, with suitably chosen parameters, either mirror-like behaviour, transmission or capture, or even decay into one or two solitons.

\section{Concluding remarks}

Apart from increasing the store of examples of models able to support an \lq integrable' defect, this investigation was also motivated by refs\cite{heimberg2005,lautrup2011, appali2012} where the wave equation
 \begin{equation}\label{deformedboussinesq}
 \frac{1}{c^2}\, u_{tt}=\left(u_{x}+\lambda (u_x^2) +\chi^2 u_x^3\right)_x -\kappa^2 u_{xxxx},
 \end{equation}
 associated with the Lagrangian \eqref{heimberglagrangian}, is relevant. In this context, however, there is an additional quartic nonlinear term in the Lagrangian together with an extra dimensional parameter $\chi$:
\begin{equation}\label{heimberglagrangian}
  {\cal L}=\frac{1}{2c^2}\,u_t^2 -\frac{1}{2} \,u_x^2 -\frac{\lambda}{3}\,u_x^3 -\frac{\chi^2}{4}u_x^4 -\frac{\kappa^2}{2}\,u_{xx}^2.
\end{equation}
Moreover, using the data supplied in \cite{lautrup2011} the parameter $\lambda$ is negative, $\chi$ is real, and the dimensionless combination of the two parameters $\lambda$ and $\chi$ satisfies:
\begin{equation}\label{lambdachi}
\left|\lambda/\chi\right| \ \approx 1.62.
\end{equation}

\p An interesting feature of \eqref{deformedboussinesq} is the existence of a solitary wave solution \cite{lautrup2011}. Using the conventions of the present article, this solution may be written as follows:
\begin{equation}\label{lautrupetalsoliton}
u=u_0 +\frac{3\,\kappa\, \gamma}{\lambda\,\epsilon}\,\ln\left(1-\epsilon \,\tanh\frac{\gamma}{2\kappa}(x-\nu t)\,\right),\ \epsilon=\frac{3\gamma\, \chi}{\sqrt{2}\,\lambda},\ \gamma(\nu)=\sqrt{1-\nu^2/c^2}.
\end{equation}
For this solution to remain finite for all values of $x$ and $t$, it is necessary that
\begin{equation}
|\epsilon|=\left|\frac{3\gamma(\nu) \chi}{\sqrt{2}\,\lambda}\right|<1,
\end{equation}
which implies
\begin{equation}\label{nubounds}
1>\frac{\nu^2}{c^2}>1-\frac{2\lambda^2}{9\chi^2}\approx 0.42,
\end{equation}
the last step using the approximation for $|\lambda/\chi|$ provided by eq\eqref{lambdachi}. Then, eq\eqref{nubounds} provides upper and lower bounds for the speed $\nu$ of the soliton \cite{lautrup2011}. Moreover, expanding the logarithm as a power series, the solution is approximated by
\begin{equation}
u=u_0-\frac{3\gamma\, \kappa}{\lambda}\,\tanh\left(\frac{\gamma}{2\kappa}(x-\nu t)\right)+O(\epsilon),
\end{equation}
which is exactly the Boussinesq soliton of eq\eqref{Bsoliton} perturbed by contributions proportional to $\epsilon$.

\p The total momentum and energy of the solution \eqref{lautrupetalsoliton} can also be calculated exactly and then expanded as power series in $\epsilon$. The expressions for these are:
\begin{eqnarray}\label{Plautrupetal}
&&\nonumber P=\frac{9\gamma^3 \kappa\nu}{c^2\lambda^2}\,\frac{1}{\epsilon^{3}}\,\left[\ln\left(\frac{1+\epsilon}{1-\epsilon}\right)
-2\epsilon\right]=\frac{9\gamma^3 \kappa\nu}{c^2\lambda^2} \,\sum_{k=0}^\infty\, \frac{2\,\epsilon^{2k}}{2k+3}\\ &&\phantom{P}\approx M_0\gamma^3 \nu \left(1 +\frac{3 \epsilon^2}{5}\right)+O(\epsilon^4),
\end{eqnarray}
and
\begin{eqnarray}\label{Elautrupetal}
&&\nonumber E=\frac{3\kappa \gamma^3}{\lambda^2}\,\frac{1}{\epsilon^5}\left[\gamma^2 \epsilon^3 +\frac{3}{2}\left(\ln\left(\frac{1+\epsilon}{1-\epsilon}\right)-2\epsilon\right)
\left(\left(1+\frac{v^2}{c^2}\right)\epsilon^2 -\gamma^2\right)\right]\\
\nonumber &&\phantom{E}=M_0 c^2\gamma^3\left[\sum_{k=0}^\infty\frac{3\epsilon^{2k}}{(2k+3)(2k+5)}\left(1+2(k+2) \, \frac{v^2}{c^2}\right)\right]\\
&&\phantom{E}\approx \frac{M_0 c^2 \gamma^3}{5}\,\left( \left(1+\frac{4v^2}{c^2}\right) +\frac{3\epsilon^2}{7}\left(1+\frac{6v^2}{c^2}\right)\right)+O(\epsilon^4),
\end{eqnarray}
where $M_0$ is the mass parameter introduced in \eqref{EPsoliton}. As expected, both expressions reflect the fact that the solitary wave \eqref{lautrupetalsoliton} is a perturbation of the Boussinesq soliton. However, for the energy and momentum, the perturbations away from the Boussinesq soliton energy and momentum are second order in the parameter $\epsilon$.

\p It will be interesting to see to what extent the solution \eqref{lautrupetalsoliton} will be able to navigate a defect, given the perturbed Boussinesq equation \eqref{deformedboussinesq} is not itself integrable. An analysis of that will be presented in the future. It would also be interesting to explore the possible existence of deformed merging or splitting soliton solutions to eq\eqref{deformedboussinesq}.

\p In this article nothing has been said about the quantum version of the Boussinesq model, which has received attention in the past owing to its relationship to $W_3$ conformal field theory (see, for example, refs\cite{bhk2002, mr2020}). It would be interesting to investigate the existence and properties of a \lq transmission/reflection' matrix to represent the defect. This would have to be compatible with the Boussinesq S-matrix, which itself deserves further study, especially given the existence of classical solutions that do not preserve the total number of solitons. The idea would be to parallel the developments made in this direction within the sine-Gordon model and affine Toda field theories (see, for example, \cite{kl1999,bcz2005,cz2007, cz2009a, cz2010, w2010}). Not all the features noted here might be present in the Boussinesq quantum field theory but some might be.

\p There is more work to do to establish the integrability of the Boussinesq model with one or more defects. In other integrable models, such as the sine-Gordon, KdV and nonlinear Schr\"odinger examples, it has turned out to be possible to show that the full integrability is not destroyed by introducing momentum-preserving defects (for example, see references \cite{dms1994, hk2007,cz2005,cz2009,ad2012nls,z2014}). It is expected that similar arguments will apply also in the case of the Boussinesq model though they have not been explored here. If for some reason a future analysis shows that the Boussinesq theory provides a counter example to this expectation then that too would be a worthwhile result.

\p Finally, as remarked in the previous section, the Boussinesq equation appears to be the first example where an integrable defect with suitable choices of parameters  allows reflection besides the transmission or capture of solitons.  More remarkably, it is also possible to set up a defect with a store of momentum (and energy) that is subsequently partially released by the emission of one or two solitons. In other models, such as sine-Gordon, or nonlinear Schr\"odinger, this simply does not happen: integrable defects allow transmission and capture, while integrable boundaries imply perfect reflection. In the cases examined here the behaviour is more typical, for example, of a linear wave equation with a point impurity represented by a delta-function potential.  Asking why the additional phenomena are possible for the Boussinesq model but not for others will surely repay further study. The behaviour of solitons in the presence of Boussinesq defects also offers alternative means to control solitons,  with the aim of using them along the lines that have been suggested in ref\cite{cz2004}. Given the inspiration for this study provided by refs\cite{heimberg2005,lautrup2011, appali2012}, one can speculate that nature might already be using properties of \lq defects', similar to those described in this article, to facilitate the control and manipulation of solitons travelling along nerve fibres.

\end{document}